\documentclass[aps,twocolumn,superscriptaddress,groupedaddress]{revtex4}  
\usepackage{graphicx}  
\usepackage{dcolumn}   
\usepackage{bm}        
\usepackage{amssymb}   
\usepackage{amsmath}
\usepackage{mathtools}
\usepackage{appendix} 

\usepackage{color}

\begin{document}



\title{Fast neutrino cooling of nuclear pasta in neutron stars: molecular dynamics simulations} 
\author{Zidu Lin }
\affiliation{Department of Physics, Arizona State University, \\ 450 E. Tyler Mall, Tempe, AZ 85287-1504 USA}
\author{Matthew E. Caplan }
\affiliation{Department of Physics, Illinois Sate University, Normal, IL 61790, USA}
\author{Charles J. Horowitz }
\affiliation{Center for the Exploration of Energy and Matter and Department of Physics, Indiana University, Bloomington, IN 47405, USA}
\author{Cecilia Lunardini }
\affiliation{Department of Physics, Arizona State University, \\ 450 E. Tyler Mall, Tempe, AZ 85287-1504 USA}

\date{\today}

\begin{abstract}
The direct Urca 
process of rapid neutrino emission can occur in nonuniform nuclear pasta phases that are expected in the inner crust of neutron stars. 
Here, the periodic potential for a nucleon in nuclear pasta allows momentum conservation to be satisfied for direct Urca reactions. We improve on earlier work by modeling a rich variety of pasta phases (gnocchi, waffle, lasagna, and anti-spaghetti) with large-scale molecular dynamics simulations.
We find that the neutrino luminosity due to direct Urca reactions in nuclear pasta can be 3 to 4 orders of magnitude larger than that from the modified Urca process in the NS core.  Thus neutrino radiation from pasta could dominate radiation from the core and this could significantly impact the cooling of neutron stars. 
\end{abstract}

\maketitle

\section{Introduction}


Neutron stars (NS) cool primarily by neutrino emission from their dense interiors \cite{annurev.astro.42.053102.134013,Page_2004}.   Therefore, X-ray observations of NS surface temperatures  can provide insight into exotic high density phases that may be present.  Many neutron stars are thought to cool relatively slowly by the modified Urca process where two correlated nucleons undergo a cycle of beta decay followed by electron capture that radiates neutrino anti-neutrino pairs \cite{Page_2004}. Two nucleons are needed in order to conserve both momentum and energy during the weak interactions.  However, this restricts the available phase space and reduces the neutrino emissivity $Q_{\mathrm{mUrca}}$ of the modified Urca process.

Alternatively, if the proton fraction in dense matter is very high, above a critical value of $Y_{P}^{C}=0.11-0.15$ \cite{Lattimer:1991ib}, it is possible for a single neutron to beta decay and conserve both momentum and energy.  This leads to the direct Urca process that has a much higher neutrino emissivity \cite{Lattimer:1991ib}.   The high proton fraction necessary for direct Urca can only be achieved at sufficiently high densities for those equations of state with large symmetry energies \cite{Lattimer:1991ib}. Until recently, the direct Urca process has been generally believed to occur in the inner cores of very high mass neutron stars \cite{page1992cooling}, where the density is large enough so that $Y_{P}^{C}$ 
is achieved. If direct Urca is allowed, it can serve as the most important cooling channel and its presence can be tested by X-ray observations of NS thermal radiations.  For example, recently the neutron star in MXB 1659-29 was observed to have a very low surface temperature, despite large accretion heating.  This strongly suggests enhanced neutrino cooling from a direct Urca or similar process \cite{Brown:2017gxd}. 
The original direct Urca process occurs in the inner cores of massive NSs.   More recently a number of direct Urca like processes that take place at lower densities are being explored. Schatz \emph{et al.}  discuss a possible Urca cycle where a large amount of energy is emitted as neutrinos, mainly from the nuclei with odd mass-number A 
in the outer crust of a NS, which undergoes first beta decay and then electron capture \cite{Schatz:2013xea}.  Here nuclear recoil allows the conservation of both momentum and energy and neutrino radiation from this cycle could rapidly cool a layer of the crust.



Nonuniform phases of nuclear matter may allow another way to conserve both momentum and energy for the weak interactions.  At just below nuclear density, competition between coulomb repulsion and nuclear attraction can rearrange nuclear matter into rod-like, slab-like, 
or other complex shapes that are known as nuclear pasta \cite{PhysRevLett.50.2066, 10.1143/PTP.71.320}. Nuclear pasta is expected at the base of the NS crust, just before the transition to uniform nuclear matter in the NS core \cite{Watanabe:2011yz, RevModPhys.89.041002}.

In \cite{Gusakov:2004mj}, Gusakov \emph{et al.} showed that the direct Urca process can possibly occur in nuclear pasta.  
Due to the periodic potential created by the inhomogeneous density distributions of nuclear pasta, nucleons in the inner crust would acquire large quasi-momenta, and in this way satisfy the momentum conservation required by the direct Urca process. Gusakov \emph{et al}.
\cite{Gusakov:2004mj} use a liquid drop model by K. Oyamatsu \cite{Oyamatsu:1993zz} to describe the pasta and focus on two high density pasta phases (inverted cylinder and inverted sphere) 
when calculating the neutrino emissivity. Gusakov \emph{et al.} found that the neutrino emissivity due to the direct Urca process in a layer of nuclear pasta can be 2 orders of magnitude stronger than the modified Urca process (although still about 5 orders of magnitude weaker than the Urca process in the neutron star core).  Thus, in a neutron star where the central density is too low to support the direct Urca reaction in the core, this neutrino emission reaction in the NS crust can profoundly affect NS cooling. 

In addition, Urca emission from nuclear pasta could modify the cooling of neutron star crusts in similar ways to Ref. \cite{Schatz:2013xea}. Here the surface layers could thermally decouple from the deeper regions so that X-ray bursts and other surface phenomena might be independent of the strength of deep crustal heating.

Finally, a strong Urca process could produce significant bulk viscosity, which arises from the phase shift between NS density oscillations and the restoration rate of beta equilibrium via Urca emissions \cite{Yakovlev:2018jia, Alford:2019qtm}. Such a strong bulk viscosity in the nuclear pasta layer could dominate over that in outer NS cores due to modified Urca process, and could be important for the damping of r-mode oscillations in NSs. 



In this paper, we present improved calculations of the neutrino emissivity of pasta based on large-scale molecular dynamics (MD) simulations.  These semiclassical simulations allow us to freely explore more complex nuclear pasta shapes and to directly calculate the emissivity.
The method we use in this work has been extensively used in the past to study the thermal conductivity, electrical conductivity, shear viscosity and neutrino opacity of nuclear pasta \cite{Horowitz:2004pv,Schneider:2013dwa,Horowitz:2016fpa}. We investigate the effect of four main pasta phases (gnocchi, waffle, lasagna, anti-spaghetti) observed in our MD nuclear pasta model on the direct Urca process in the inner crust. The paper is organized as follows: in Section \ref{method}, we discuss our molecular dynamics simulations of the nuclear pasta, and the physics of the direct Urca process in the pasta layer. In Section \ref{result}, we present the calculation of the neutrino emissivity, which is very sensitive to the nuclear pasta structure. We then calculate the neutrino luminosity due to direct Urca process in the pasta and compare it with the luminosity due to the modified Urca process in the NS core. Finally, we conclude in Section \ref{conclude}.   

\section{Method}\label{method}

\subsection{Direct Urca emissivity and its reduction factor in neutron star crust}\label{sec:RF}

To calculate the neutrino emissivity of the direct Urca process in neutron star crusts, we first determine the wave functions of protons and neutrons in nuclear pasta layer, which is roughly approximated using perturbation theory and is expressed as a Bloch wave function as in \cite{Gusakov:2004mj}. The nucleon wave function is written as:

\begin{equation}
    \Phi_j=\dfrac{\chi_s}{\sqrt{V}}(e^{i\textbf{p}\cdot\textbf{r}}+\sum_{\textbf{q}\neq0}C_{\textbf{q}}e^{i\textbf{p}'\cdot\textbf{r}}),
\end{equation}
where $V$ is the normalization volume, $\textbf{q}$ is the inverse lattice vector, $\textbf{p}$ is the momentum of a nucleon, $C_{\textbf{q}}=V_j(\textbf{q})/(E_{\textbf{p}}-E_{\textbf{p}'})$ and $\textbf{p}'=\textbf{p}+\textbf{q}$. Finally $V_j(\textbf{q})$ is the Fourier transformed nucleon potential in the nuclear pasta with $j=N,P$, where $N$ stands for a neutron and $P$ stands for a proton. Given the nucleon wave functions, the neutrino emissivity $Q$ is calculated similarly as in \cite{Gusakov:2004mj}, and we get:
\begin{equation}\label{Q}
    Q(T,n)=Q_0(T,n)R(n),
\end{equation}
where $Q_0$ is the direct Urca emissivity in uniform matter without the momentum conservation constraint, $T$ is temperature and $n$ is baryon number density.  Specifically, $Q_0$ is written as:
\begin{equation}
\label{q0}
\begin{split}
    Q_0(T,n)&=\dfrac{457\pi}{10080}G_F^2\cos^2\theta_C(f_V^2+3g_A^2)m_Nm_Pm_eT^6\\&\approx4\times10^{27}\left(\dfrac{n_e}{n_0}\right)^{1/3}T_9^6 \quad \mathrm{erg\,cm^{-3}s^{-1}},
\end{split}
\end{equation}
with $T_9=T/10^9$ K. In the calculations we assume that $m_{N,P}^*=m_{N,P}$, where $m_{N,P}^*$ is the effective mass of a neutron or a proton at the Fermi surface.  
The effect of the nuclear pasta structure on $Q$ is manifested in the function $R(n)$, which is: 
\begin{equation}\label{eq:R}
\begin{split}
    R(n)=&\sum_{j=N,P}\sum_{\textbf{q}}\dfrac{(m_jV_{j}(\textbf{q}))^2}{\alpha_jP_{Fj}^4}\\&\times[F(2\alpha_jD_j^{\mathrm{max}}+2\alpha_j^2)-F(2\alpha_jD_j^{\mathrm{min}}+2\alpha_j^2)]\\&\times\Theta,
\end{split}
\end{equation}
where $P_{Fj}$ is the Fermi momentum of a nucleon, $\alpha_j=q/P_{Fj}$, $D_{N\pm}=[(P_{FP}\pm P_{Fe})^2-P_{FN}^2-q^2]/2P_{FN}q$, $D_{P\pm}=[(P_{FN}\pm P_{Fe})^2-P_{FP}^2-q^2]/2P_{FP}q$, $D_j^{\mathrm{max}}=\mathrm{min}[1,D_{j+}]$, $D_j^{\mathrm{min}}=\mathrm{max}[-1,D_{j-}]$, and $F(x)=\dfrac{1}{2}\mathrm{ln}|(\sqrt{1+x}+1)/(\sqrt{1+x}-1)|-\sqrt{1+x}/x$ . Following \cite{Gusakov:2004mj}, a simplified Thomas-Fermi approximation is used in our calculation, and the Fermi momentum of a neutron and a proton is calculated as: $P_{FN}=(3\pi^2n_N)^{1/3}$ and $P_{FP}=P_{Fe}=(3\pi^2n_P)^{1/3}$. Finally, $\Theta$ is a step function: $\Theta=1$ if the momentum conservation is satisfied in the direct Urca like reactions, and $\Theta=0$ otherwise. The step function constrains the region of allowed momentum transfer $\textbf{q}$ in direct Urca reactions in nuclear pasta layer: 

\begin{equation}
 P_{FN}-P_{FP}-P_{Fe}\leq q\leq P_{FN}+P_{FP}+P_{Fe}. 
 \label{eq:momentum conservation}
\end{equation}

To determine $R$, we need to specify the baryon density $n$, the electron fraction $Y_e$ as well as the Fourier transformed nucleon potential $V_j$(\textbf{q}) in pasta phases. In this work our MD simulations are used to find the baryon density $n$ at which the nuclear pasta phases of gnocchi, lasagna, waffle, and anti-spaghetti form. 
A detailed description of the MD simulation is presented in section \ref{sec:MD}. The electron fraction $Y_e$ in the pasta layer of neutron stars should be applied in eq. (\ref{Q}). Oyamatsu  \cite{Oyamatsu:1993zz} studied the nuclear pasta at beta equilibrium, and found that the pasta forms at proton fraction $Y_P\approx0.03$. Correspondingly we calculated the function $R(n)$ at around $Y_P= 0.03$, to study the direct Urca process in realistic conditions of inner NS crust. 

The Fourier transformed nucleon potential $V_j(\textbf{q})$ is obtained directly from our numerical simulations of different pasta phases and will be described in more details in section \ref{sec:pastapotential}. Interestingly, we found high peaks of $V_j(\textbf{q})$ based on our large scale nuclear pasta simulations, which could potentially amplify the value of $R$ in eq. (\ref{eq:R}), and could give a much larger neutrino emissivity. More details about the impact from peaks of $V_j(\textbf{q})$ on the neutrino emissivity will be discussed in section. \ref{result:emissivity}.     

\subsection{Nucleon Potential in Pasta}\label{sec:pastapotential}
In section \ref{sec:RF}, we show that the effect of nuclear pasta structure on the direct Urca like process is manifested in eq. (\ref{eq:R}), where $R\propto V^2_j(\textbf{q})$. In this section we further calculate the potential energy of a nucleon $V_j$ in nuclear pasta numerically. The Indiana University semi-classical Molecular Dynamics simulation IUMD \cite{Schneider:2013dwa} is used to study the nuclear pasta structure, and the code is described with more details in section \ref{sec:MD}. In MD simulations the dynamical evolution of $N_{\mathrm{tot}}$ nucleons is simulated in a box with side length $L_{\mathrm{box}}=(N_{\mathrm{tot}}/n)^{\frac{1}{3}}$, and the structure of nuclear pasta is depicted by the time-dependent spatial distributions of nucleons in the box, which are called trajectory configurations of the nuclear pasta. In this work, the nuclear pasta phases we explored are simulated in boxes with $L_{\mathrm{box}}$ ranging from 83.4 $\mathrm{fm}$ to 150.8 $\mathrm{fm}$, and the potential energy of a nucleon $V_j$ ($j=N,P$) at $\textbf{r}_l$ is:
\begin{equation}\label{vnr}
V_j(\textbf{r}_l)=\sum_{m=1}^{N_{\mathrm{tot}}}V (l,m),
\end{equation}
where $V(l,m)$ is a semi-classical potential for a two body nucleon interaction with the spacing of nucleons being $r_{lm}=|\textbf{r}_l-\textbf{r}_m|$ (see eq. (\ref{vnucleon}) for detailed definition of $V(l,m)$), 
and $\textbf{r}_l=s d \hat{i} + t d \hat{j} + u d \hat{k}$, with $d$ being the spacing of the potential grids, $s$, $t$, $u$ being integers and $\hat{i}$, $\hat{j}$, $\hat{k}$ are orthogonal unit vectors. Consequently, we have $N_{\mathrm{grid}}=(L_{\mathrm{box}}/d)^3$ grid points on which we calculate the nucleon potential of nuclear pasta. The grid point spacing $d$ is chosen so that $d \ll L_{\mathrm{box}}$ and is approximately 2 fm, near the characteristic nucleon spacing in our model.

Given $V_j(\textbf{r}_l)$, we calculate the Fourier transformed nucleon potential $V_j(\textbf{q})$ numerically, as:
\begin{equation}
	V_j(\textbf{q})=\frac{\sum_{j=1}^{N_{\mathrm{grid}}}V_N(\textbf{r}_j)\times exp(i\textbf{q}\cdot\textbf{r}_j)d^3}{L_{\mathrm{box}}^3},
\end{equation}
where $\textbf{q}=(\frac{2\pi}{L_{\mathrm{box}}}M)\hat{i}+(\frac{2\pi}{L_{\mathrm{box}}}N)\hat{j}+(\frac{2\pi}{L_{\mathrm{box}}}O)\hat{k}$, with $M$, $N$, $O$ being integers.

We use 100 trajectory configurations  from the MD nuclear pasta simulations spaced by 1000 MD timesteps. The nucleon potential  $V_j^x(\textbf{r}_j)$ of each configuration $x$ is calculated per eq. (\ref{vnr}), and is 
averaged by $V_j(\textbf{r}_i)=\sum_{x=1}^{100} V_N^x(\textbf{r}_i)/100$. 

\subsection{Molecular Dynamics of Nuclear Pasta}\label{sec:MD}


We use the Indiana University Molecular Dynamics code (IUMD) to simulate nuclear pasta, as in past work \cite{PhysRevC.69.045804,Schneider:2013dwa,PhysRevC.91.065802,PhysRevLett.114.031102,PhysRevC.93.065806,PhysRevC.90.055805,RevModPhys.89.041002,PhysRevC.98.055801,PhysRevLett.121.132701,caplan2020thermal,PhysRevC.94.055801}. For completeness, we include a brief review here. IUMD uses a semi-classical potential $V(l,m)$ for a two body nucleon interaction, which is:  
\begin{equation}\label{vnucleon}
V(l,m)=ae^{-r_{lm}^2/\Lambda}+[b+c\tau_z(l)\tau_z(m)]e^{-r_{lm}^2/2\Lambda}+V_c(l,m).
\end{equation}
Here $a=110$ MeV, $b=-26$ MeV, $c=24$ MeV, 
$\Lambda=1.25$ $\mathrm{fm}^2$, and 
\begin{equation}
V_c(l,m)=\frac{e^2}{r_{lm}}e^{-r_{lm}/\lambda}\tau_P(l)\tau_P(m)
\end{equation}
is the Coulomb repulsion between protons. We set $\lambda=10$ fm as the Coulomb screening length, $\tau_z=1$ for proton and $\tau_z=-1$ for neutron, and $\tau_p=\frac{1+\tau_z}{2}=1$. Note that $r_{ij}=\sqrt{[x_i-x_j]^2+[y_i-y_j]^2+[z_i-z_j]^2}$ where the periodic distance (given by $[l]=\mathrm{Min}(|l|,L-|l|)$) is used. 
All simulations described in this work use periodic boundary conditions in a cubic box, with side length $L$. 
All simulations are isothermal and at constant density with an MD timestep of 2 fm/c.  

This two-body interaction is simple and the nuclear attraction is short ranged, allowing us to efficiently simulate hundreds of thousands of nucleons \cite{PhysRevC.98.055801,PhysRevC.98.055801,caplan2020thermal,RevModPhys.89.041002}. The geometric pasta phase can be specified by three thermodynamic parameters: the nucleon number density $n$, temperature $T$, and the proton (electron) fraction $Y_e$, though hysteresis effects and formation history can be relevant for determining the exact structure of large volumes of pasta \cite{RevModPhys.89.041002,Schneider:2013dwa}. This model has now been used extensively to study the phases and structure of nuclear pasta. It is known to form a variety of phases similar to diblock copolymers including gnocchi (spheres), lasagna (planar or lam), waffles (perforated lam), and antispaghetti (uniform matter with cylindrical holes), which will be the subject of this work \cite{RevModPhys.89.041002,Schneider:2013dwa,PhysRevC.98.055801,PhysRevC.90.055805}. Our model, having finite temperature, also exhibits a large variety of additional phases and `defects' as well, such as helicoids that connect lasagna (structurally identical to Terasaki ramps), and may buckle over large lengths and disrupt long range order \cite{PhysRevC.94.055801,PhysRevC.98.055801,caplan2020thermal,PhysRevLett.121.132701}. This work is therefore not confined to the unit cell; our MD model allows us to study both the simple idealized cases and phases with long range disorder self consistently, which is not possible with fully quantum mechanical simulations which are limited to small numbers of particles \cite{Fattoyev:2017zhb,PhysRevC.87.055805}. 

We address the robustness of our semi-classical model for this problem. Past work with this model has focused on the parameter space near $T=1$ MeV, nucleon densities between $n=$0.01 and 0.12 fm$^{-3}$, and electron (proton) fractions between $Y_e$ = 0.3 and 0.5 because this is the parameter range for which our model produces pasta \cite{RevModPhys.89.041002}. At significantly higher temperature the 
nucleons dissolves into a gas, while at temperatures near 0.5 MeV the nucleons crystallize and become locked into a lattice. Similarly, at lower proton fractions our model forms a gas of nucleons \cite{caplan2020thermal}. Therefore, we are confined to this parameter range to study pasta when using IUMD simulations, although the proton fraction range applied in the simulations is higher than expected in inner crust of NSs. Nevertheless, our pasta model is still consistent with mean field and fully quantum mechanical simulations which produce all the same pasta phases we observe at similar densities \cite{Stone:2009rp,Fattoyev:2017zhb}. We note that at very low proton fractions of $Y_e=0.05$ and $Y_e=0.1$, which are close to the beta equilibrium conditions, a large scale quantum simulation of pasta phases \cite{Fattoyev:2017zhb} gives similar nuclear pasta structures when comparing them with results from IUMD. The pasta phases may have many minima in their energy landscape separated by large tunneling barriers, and so configurations which are stable on MD timescales may not be true ground states. Nevertheless, initial and final configurations are generally equivalent to each other in all simulations reported in this work so these configurations are stable on MD timescales and furthermore we do not observe any trend in total simulation energy. 

As one of our primary goals in this work is to calculate the temperature independent function $R$ in eq. (\ref{Q}), which controls the magnitude of neutrino emissivity relating to pasta structures, the exact thermodynamic parameters of our pasta simulations do not heavily bias our results; we use them to generate structural conditions which contain sufficiently large numbers of nucleons to be in classical limit.

\section{Results}\label{result}

\subsection{Molecular Dynamics Simulations}\label{result:md}

\begin{table*}[t] \caption{Summary of molecular dynamics configurations studied in this work. We include nucleon number density $n$, temperature $T$, number of nucleons $N_{\mathrm{tot}}$, proton (electron) fraction $Y_e$, and their source in the literature.}\label{tab:md}
\begin{ruledtabular}
\begin{tabular}{llllll}
Identifier & $n$ (fm$^{-3}$) & $T$ (MeV) & $N_{\mathrm{tot}}$ & $Y_e$ & Source \\
\hline
G1 & 0.015000 & 1.0 & 51200 & 0.3 & Refs. \cite{2017PhDT........80C,caplanPD} \\
G2 & 0.014951 & 1.0 & 51200 & 0.4 & Refs. \cite{2017PhDT........80C,caplanPD}\\
L1 & 0.050000 & 1 & 102400 & 0.4 & Refs. \cite{PhysRevC.93.065806,PhysRevLett.121.132701} \\
L2 & 0.050000 & 1.2 & 102400 & 0.5 & Present work, derivative of L1 and Ref. \cite{caplan2020thermal} \\
L3 & 0.050000 & 1 & 204800 & 0.4 & Present work \\
L4 & 0.050007 & 0.8 & 204800 & 0.4 & Present work \\
W1 & 0.050000 & 1 & 102400 & 0.3 & Present work, derivative of L1 and Ref. \cite{caplanPD} \\
W2 & 0.05 & 1.6 & 102400 & 0.4 & Present work, derivative of Ref. \cite{PhysRevLett.121.132701} \\
W3 & 0.05 & 1 & 204800 & 0.3 & Present work, derivative of Ref. \cite{PhysRevC.93.065806} \\
W4 & 0.050007 & 0.8 & 204800 & 0.3 & Present work, derivative of L4 \\
AS1 & 0.0882 & 0.8 & 51200 & 0.3 & Present work\\
AS2 & 0.0882 & 0.8 & 51200 & 0.4 & Refs. \cite{2017PhDT........80C,caplanPD} \\
\end{tabular}
\end{ruledtabular}
\end{table*}

We study 12 MD simulations of nuclear pasta in this work, two gnocchi, four lasagna, four waffles, and two antispaghetti. We give each an identifier for readability, such as G1 and G2 for the gnocchi simulations, etc. Initial conditions for our MD simulations are assembled from or derived from our body of past work and archival data, though a few new configurations were generated for this work. The preparation of these simulations are briefly described in the following, while a summary of the molecular dynamics conditions is included in Tab. \ref{tab:md}.

The initial conditions for these simulations were all evolved 
for at least $10^6$ MD timesteps prior to collecting data to guarantee they were dynamically equilibrated. 
For consistency, all configurations used to calculate $R$ were generated from equilibrium MD simulations specifically for this work.

\begin{figure*}[htp]
	\centering
	\includegraphics[width=0.49\textwidth]{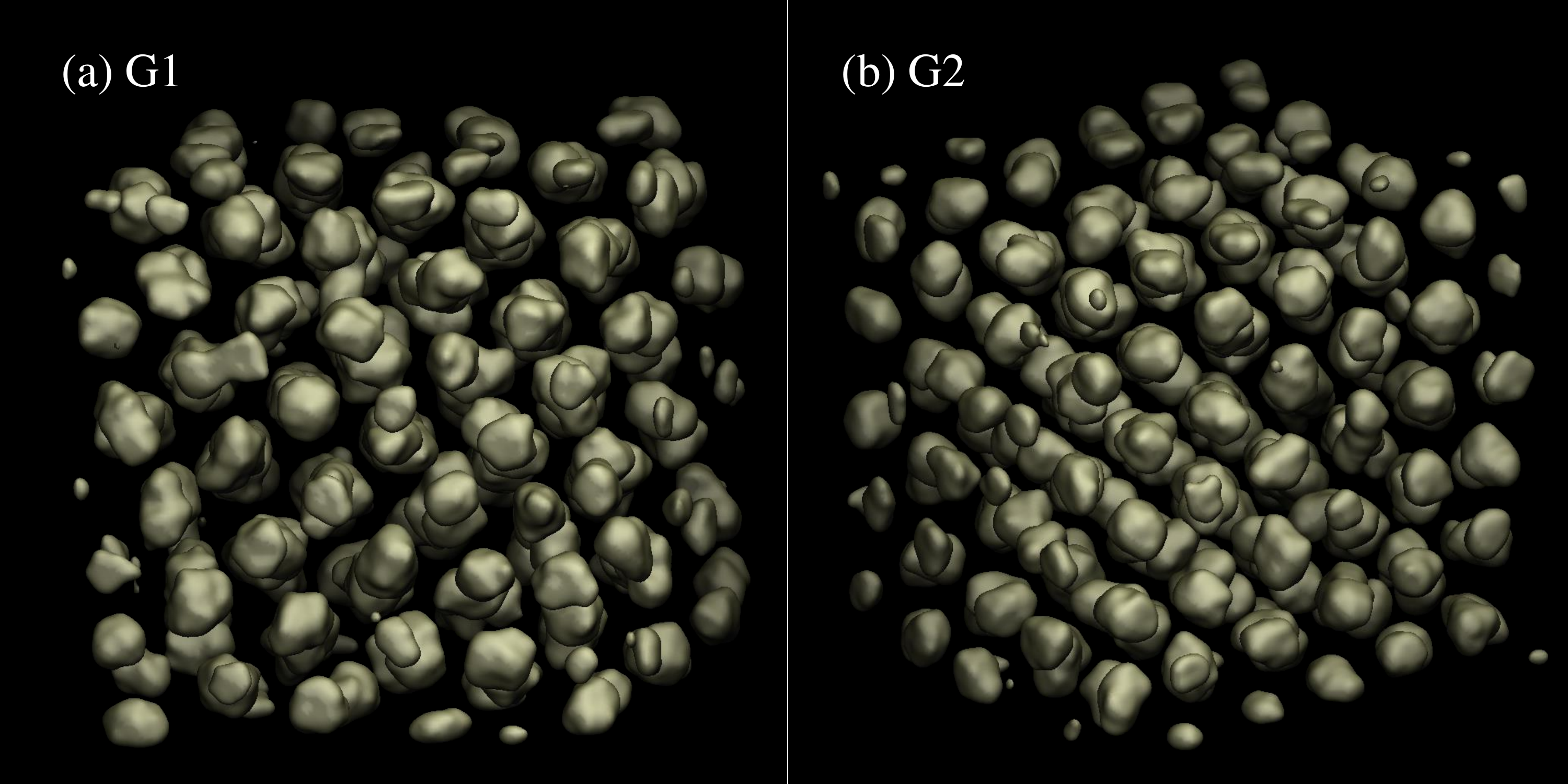}
	\includegraphics[width=0.49\textwidth]{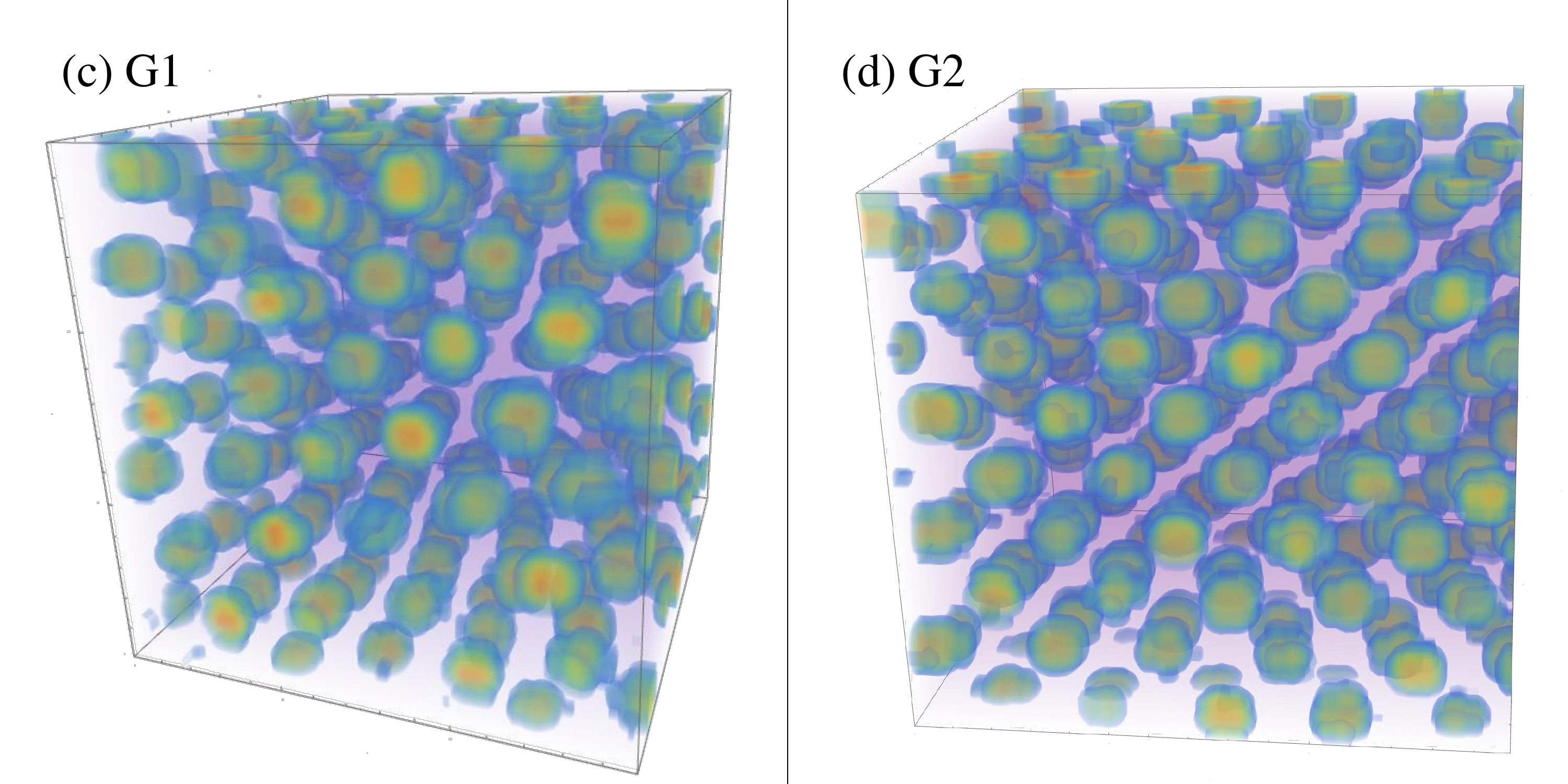}
	\caption{ (Color online) Neutron density distributions and potential energy distributions for the gnocchi phase. Panel (a) and (b) represent neutron density distribution in gnocchi simulated with 51200 nucleons at $Y_e=0.3$ and at $Y_e=0.4$ respectively. In panel (c) and (d) we show the potential energy distribution of a neutron from 0 (deep blue) to 50 (red) MeV in the gnocchi.}
	\label{fig:vrball}
\end{figure*}

G1 and G2 were taken from past studies (ref. \cite{2017PhDT........80C}) which considered phases of nuclear pasta at different proton fractions and are shown in Fig. \ref{fig:vrball}. In that work, high density matter was expanded by incrementally increasing the box size after each timestep. This results in much more regularly distributed gnocchi than in simulations equilibrated from random, as they fission from large structures generally more symmetrically. A clear body-centered-cubic (BCC) lattice  is visible, with nuclear separations comparable to nuclear radii. 

\begin{figure*}[htp]
	\centering
		\includegraphics[width=0.49\textwidth]{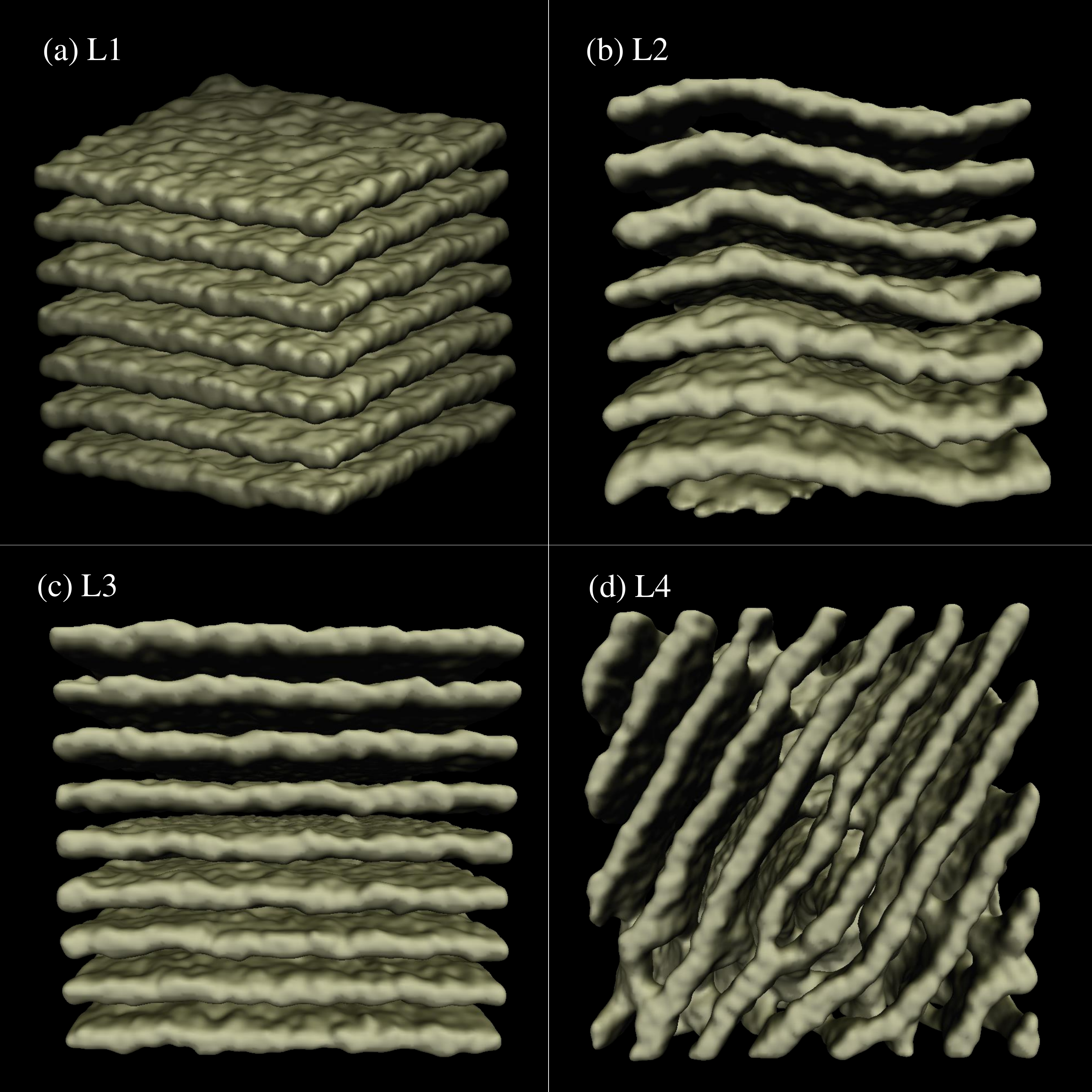}
	\includegraphics[width=0.49\textwidth]{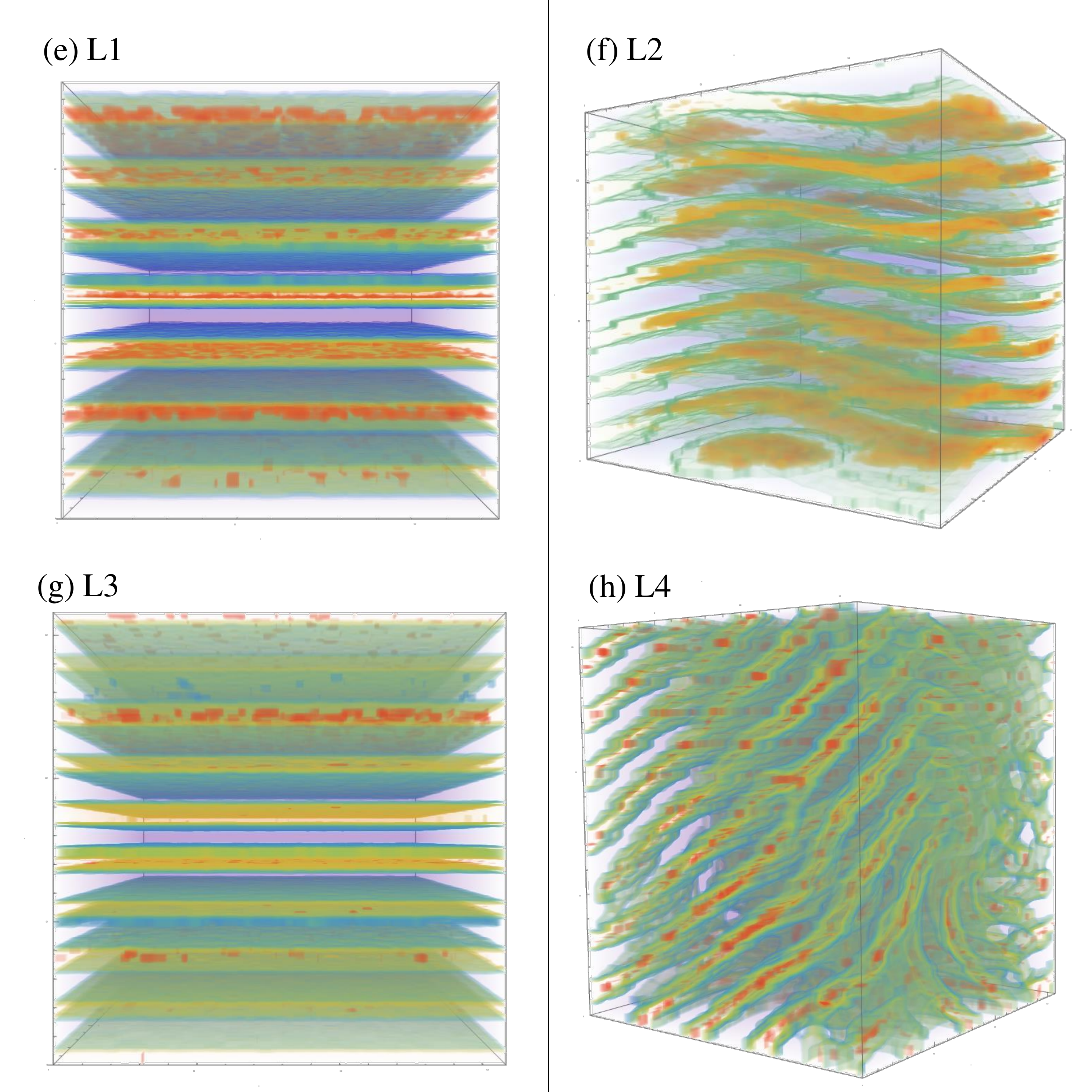}
	\caption{ Results are shown for density and real space potential energy distribution of a neutron, due to lasagna structure. Panel (a), (b), (c) and (d) represent neutron density distribution in the lasagna simulated with 102400 nucleons at Ye=0.4, 102400 nucleons at Ye=0.5, 204800 nucleons at Ye=0.4 and 204800 nucleons at Ye=0.4 but not aligned with the box surface respectively. Also, in panel (e), (f), (g) and (h) we show the potential energy distribution of a neutron from 0 (deep blue) to 50 (red) MeV in the lasagna. }
	\label{fig:vrlasagna}
\end{figure*}

Simulations of lasagna can be seen in Fig. \ref{fig:vrlasagna}. L1 and L2 were likewise taken from past work (ref. \cite{PhysRevC.93.065806,PhysRevLett.121.132701,caplan2020thermal} and allow us to study finite size effects and how the orientation of pasta within the simulation volume may affect our calculations. L1 was prepared by including a sinusoidal external potential during a brief initial simulation, while L2 is generated from L1 by random switching neutrons for protons, resulting in plates with spontaneous splay at a higher $Y_e$. L3, instead, is unique to this work, though prepared similarly to L1. L4, while having similar parameters to our other simulations of lasagna, was prepared by simulating at the slightly lower temperature of 0.8 MeV.  At this temperature many defects are frozen in, including helicoids and buckles which present a sort of `fingerprint' defect. This allows us to compare more idealized plates to a structure without long ranged order. L4 is long lived in MD.

\begin{figure*}[htp]
	\centering
		\includegraphics[width=0.49\textwidth]{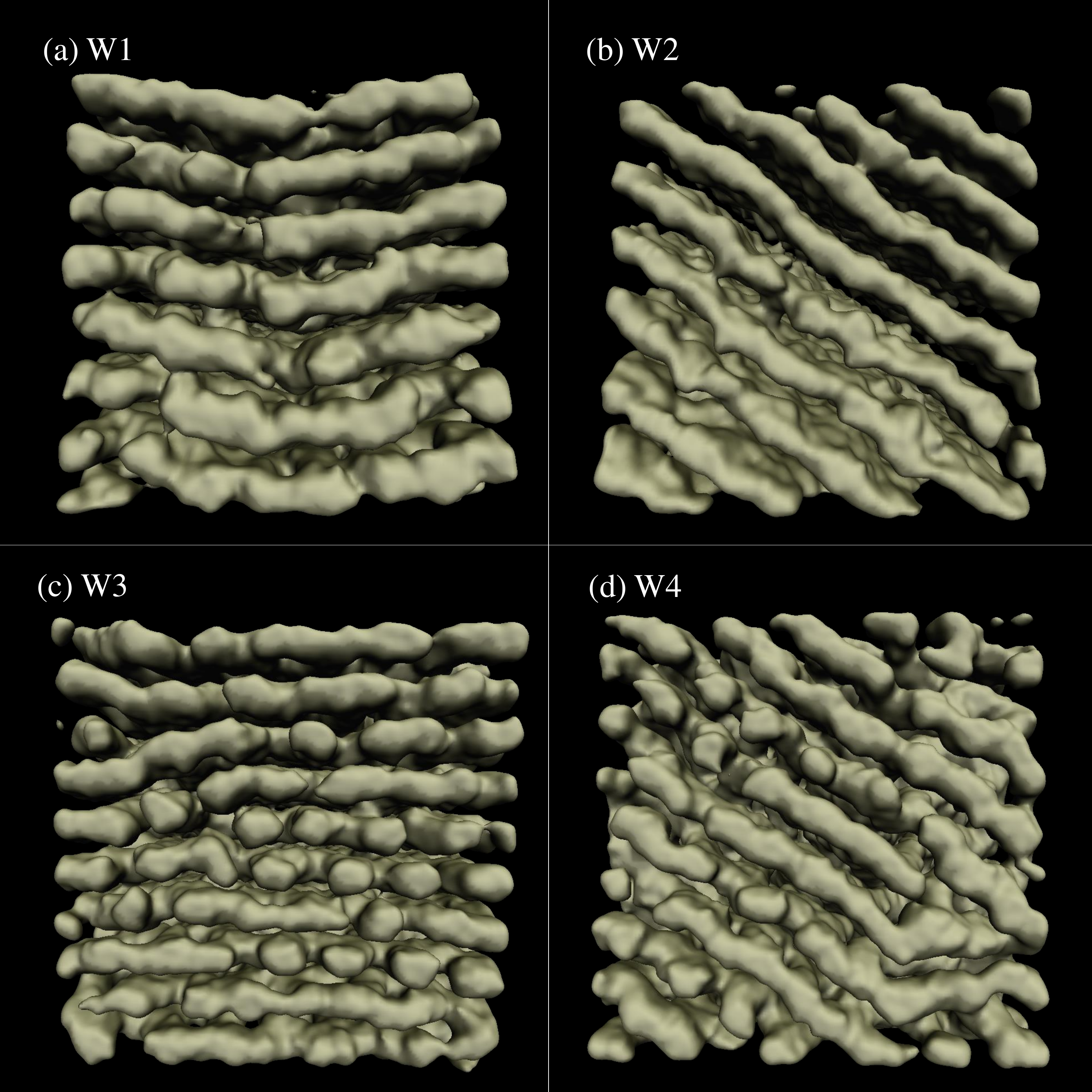}
	\includegraphics[width=0.49\textwidth]{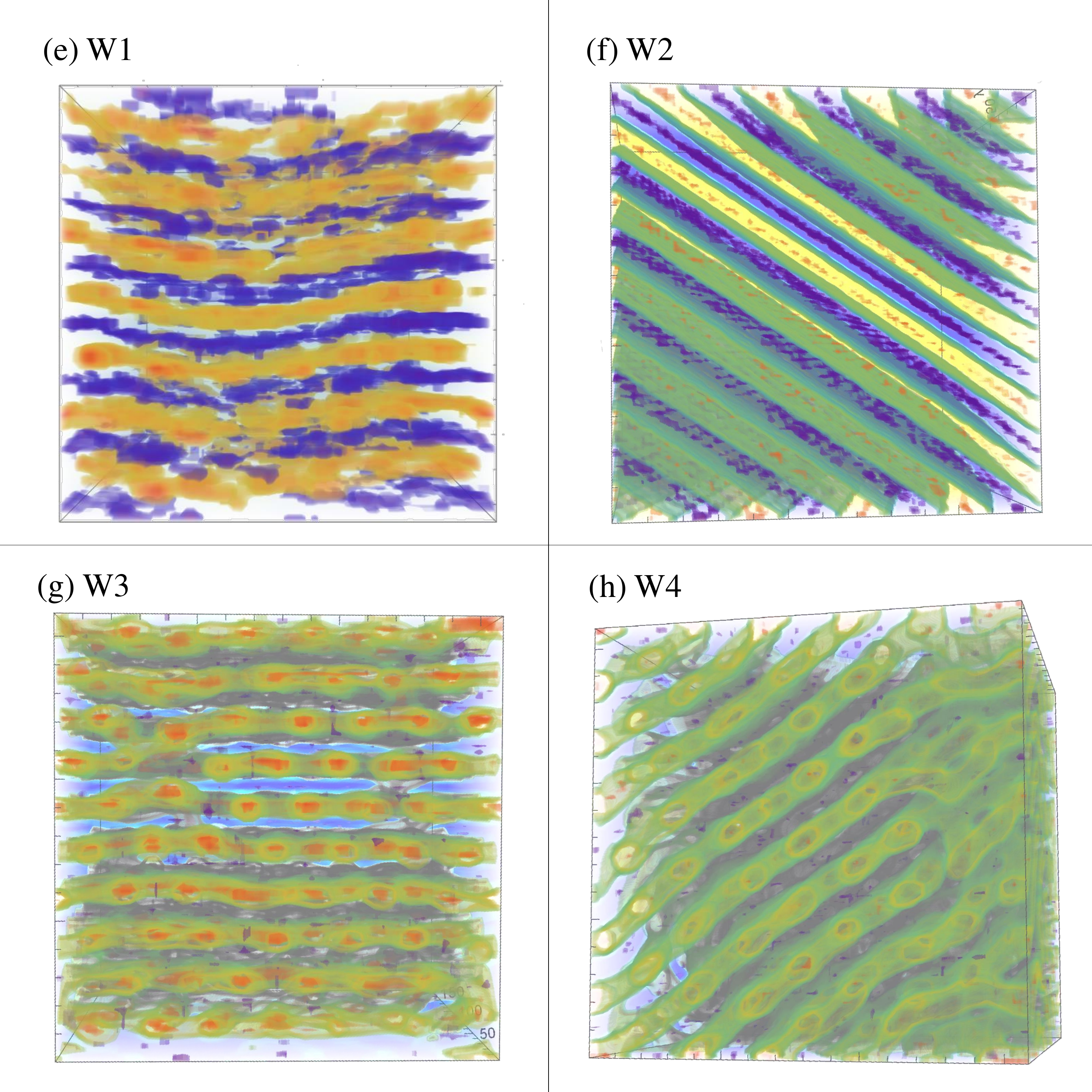}
	\caption{ Results are shown for density and real space potential energy distribution of a neutron, due to waffle structure. Panel (a), (b), (c) and (d) represent neutron density distribution in the in 'waffle' with 102400 nucleons at Ye=0.3, 102400 nucleons at Ye=0.4, 204800 nucleons at Ye=0.3 and 204800 nucleons at Ye=0.3 but not aligned with the box surface respectively. Also, in panel (e), (f), (g) and (h) we show the potential energy distribution of a neutron from 0 (deep blue) to 50 (red) MeV in the waffle.}
	\label{fig:vrwaffle}
\end{figure*}

Our waffle configurations, which are similar to lasagna but with holes perforating the plates, are shown in Fig. \ref{fig:vrwaffle}. W1 is a trivial variation of L1, obtained by reducing $Y_e$. W2 is similarly produced from past work (ref. \cite{PhysRevLett.121.132701}), and allows us to study how the orientation of the plates in the simulation volume may affect our calculations of the reduction factor. The plates in W2, however, do not show a regular lattice of holes like W1, the higher temperature results in many short-lived holes as thermal fluctuations of the pasta surface. W3 is similar to L3 and is a variation on past work (ref. \cite{PhysRevC.93.065806}), allowing us to resolve finite size effects, while W4 is obtained from L4 by reducing the proton fraction from $Y_e=0.4$ to $Y_e=0.3$.

\begin{figure*}[htp]
	\centering
	\includegraphics[width=0.49\textwidth]{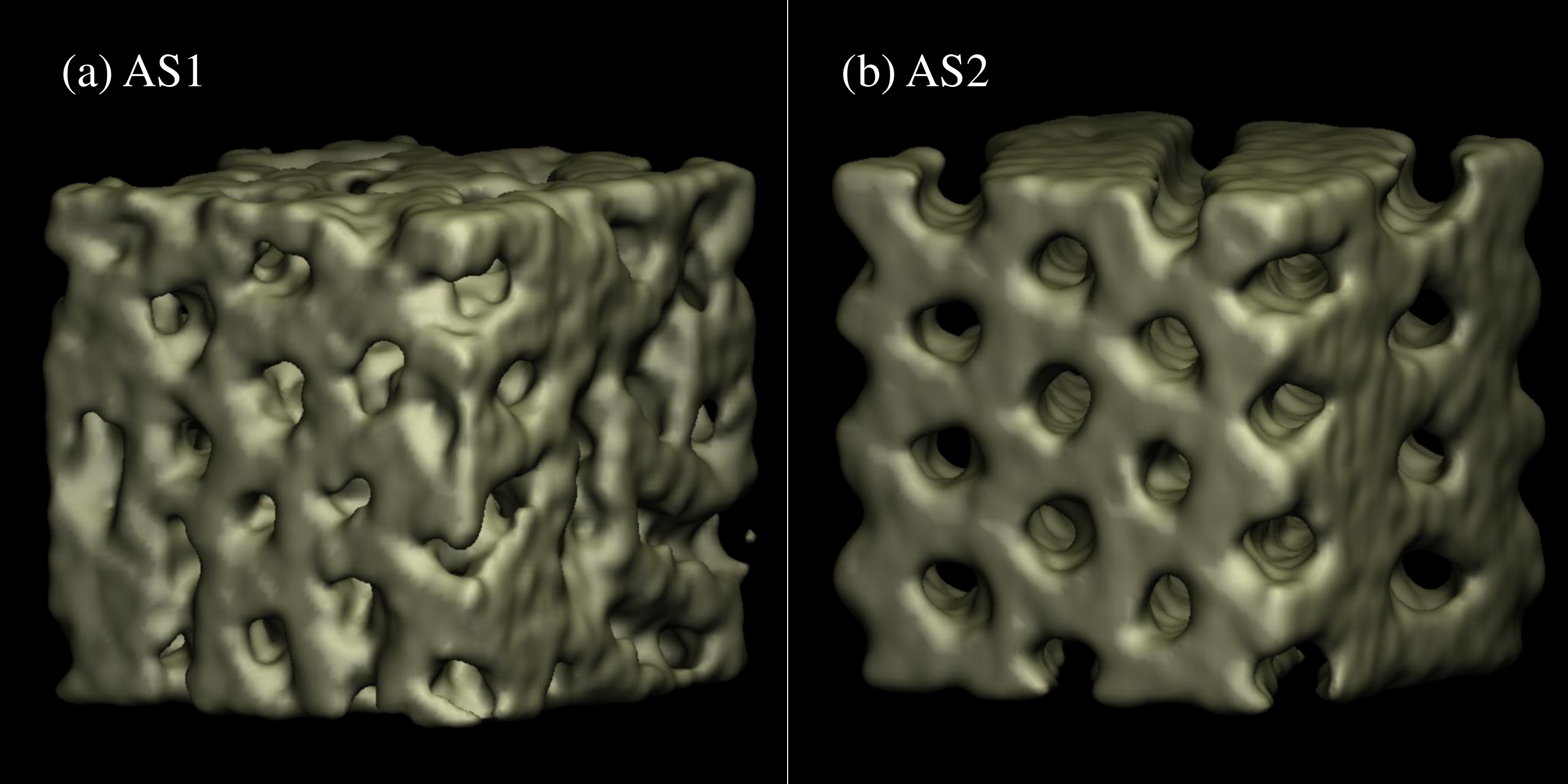}
	\includegraphics[width=0.49\textwidth]{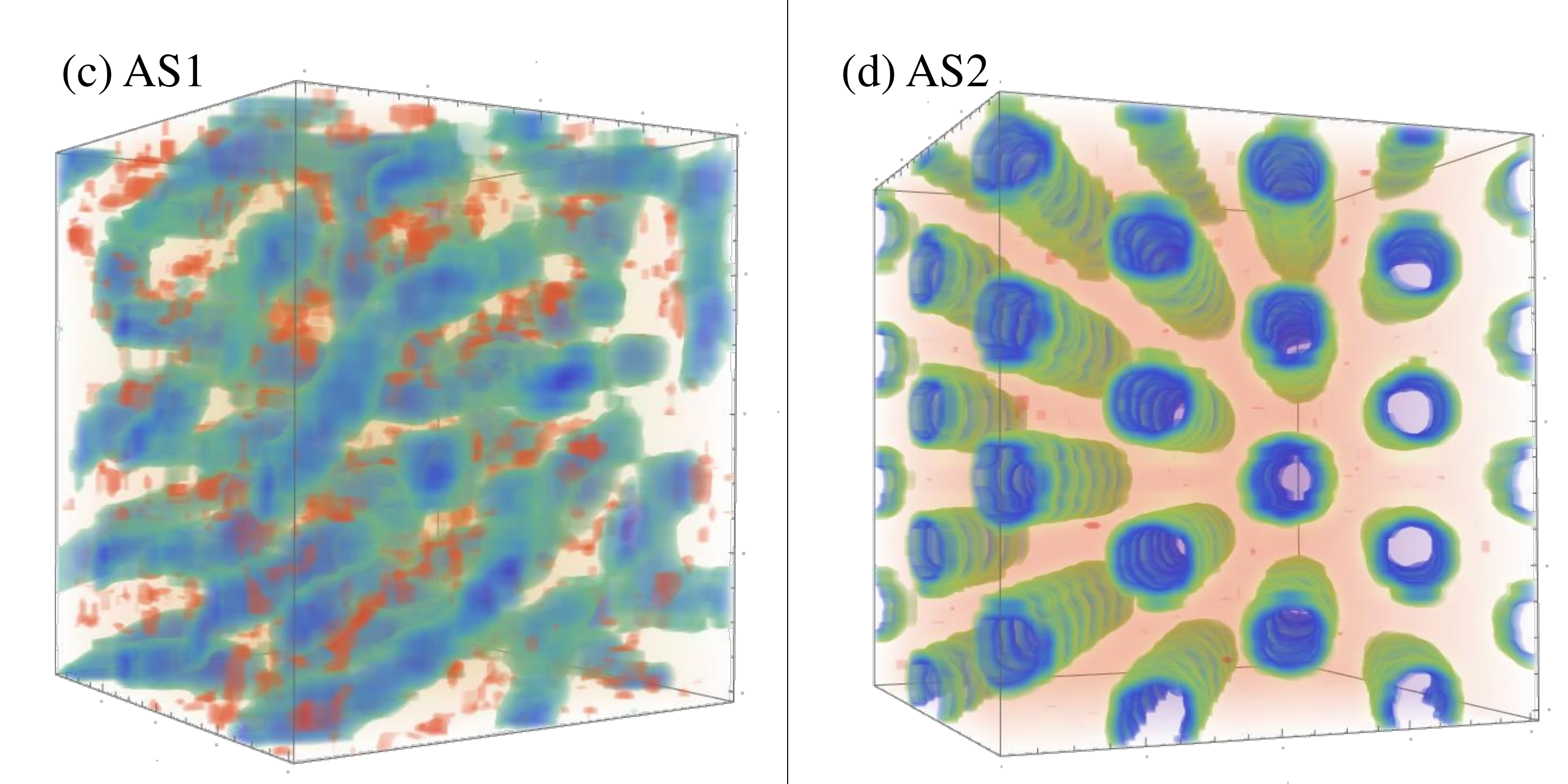}
	\caption{ Results are shown for neutron density distribution due to anti-spaghetti structure. Panel (a) and (b) represent neutron density distribution in anti-spaghetti simulated with 51200 nucleons at Ye=0.3 and 51200 nucleons at Ye=0.4 respectively. Also, Results are shown for potential energy distribution of a neutron, due to anti-spaghetti structure. In panel (c) and (d) we show the potential energy distribution of a neutron from 0 (deep blue) to 50 (red) MeV in anti-spaghetti.}
	\label{fig:vrasp}
\end{figure*}

Lastly, our antispaghetti configurations are shown in Fig. \ref{fig:vrasp}. AS1 and AS2 are obtained similarly to G1 and G2, having been taken from work which expanded dense initial conditions \cite{2017PhDT........80C}. In AS1 we resolve the low symmetry in the tunnel system, where the tunnels bend and in a few locations connect via three-way junctions.   AS1 is effectively a disordered form of AS2, which shows a high symmetry hexagonal packing of antispaghetti tunnels. 

\subsection{Nucleon potential in real space and momentum space}\label{result:vrvq}

First, we discuss the relationship between the nucleon number density distributions and nucleon potential energy distributions in nuclear pasta. Since the nuclear pasta is neutron-rich and the distributions of protons have similar structure as those of neutrons, we choose to only show the neutron density and potential distributions in Fig. \ref{fig:vrball}-\ref{fig:vrasp}. Interestingly, the potential energy distributions of nuclear pasta exhibit similar non-uniform characteristics when they are compared to the number density distributions, due to the short-range nature of the nuclear force. In this way, one might expect that the structural information of nuclear pasta will be imprinted on its Fourier transformed nucleon potential $V(\textbf{q})$ and on the magnitude of neutrino emissivity in direct Urca process (see Eq. (\ref{eq:R})).

\begin{figure*}[htp]
	\centering
	\includegraphics[width=0.45\textwidth]{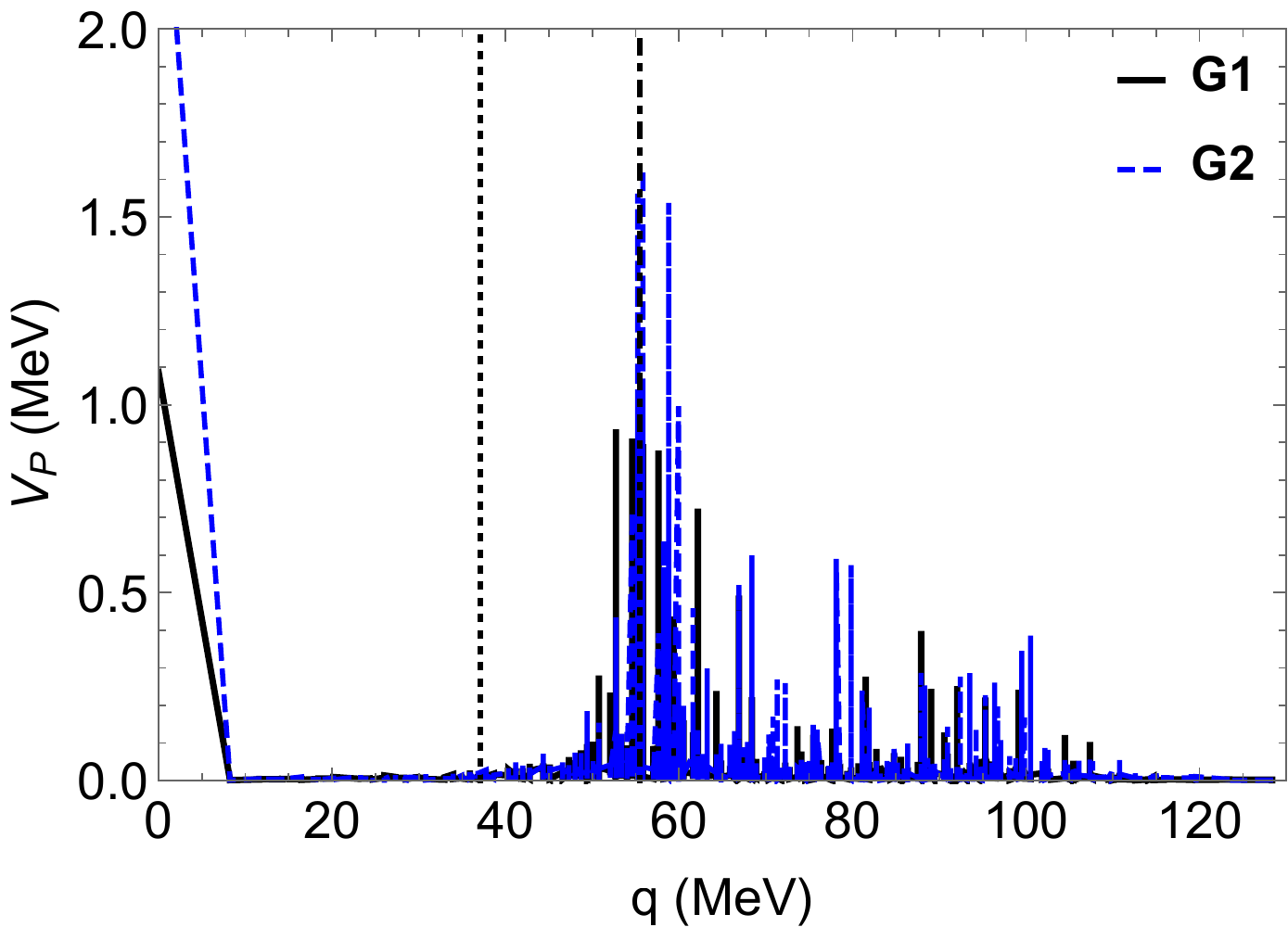}
	\includegraphics[width=0.45\textwidth]{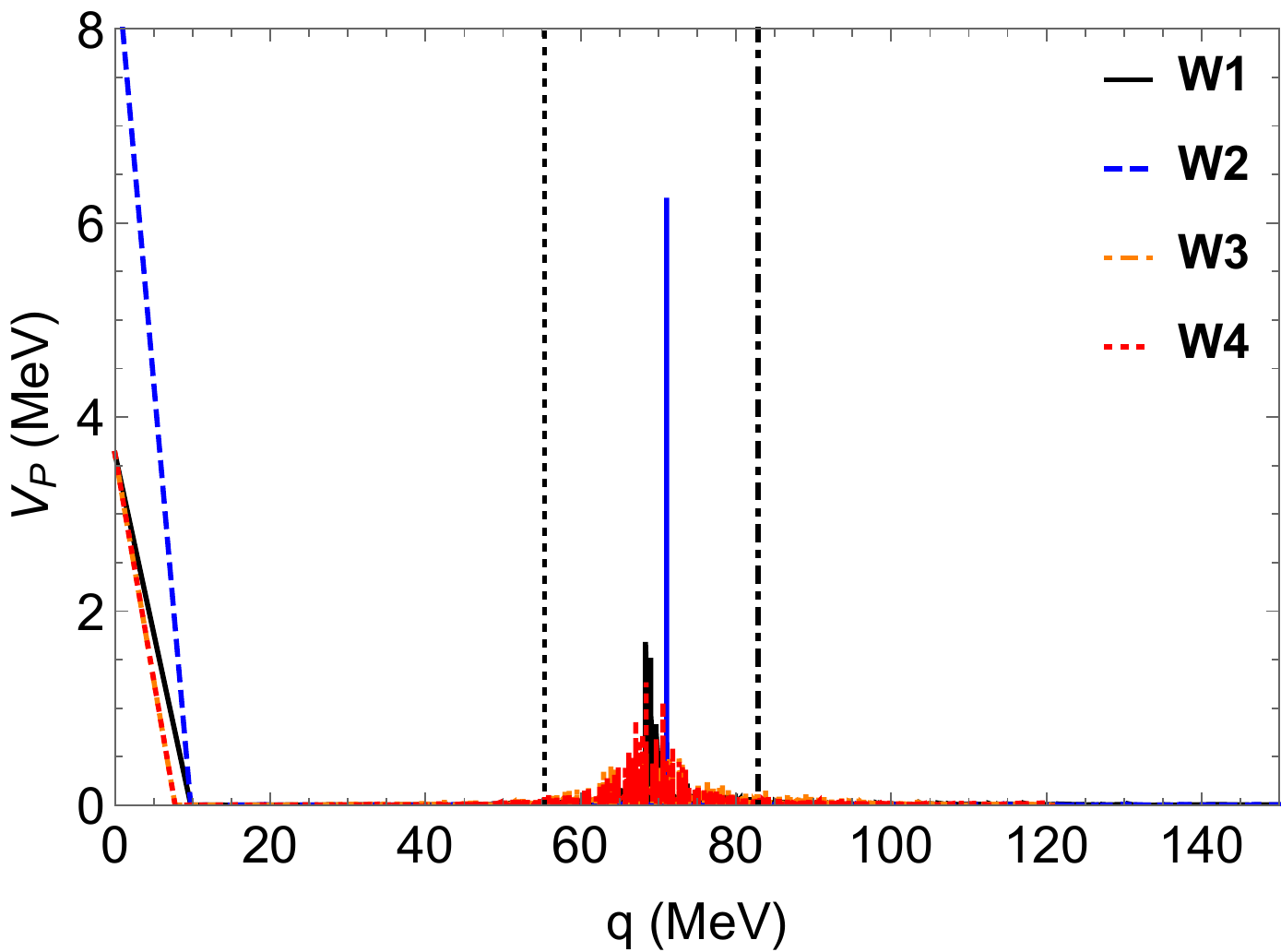}
	\includegraphics[width=0.45\textwidth]{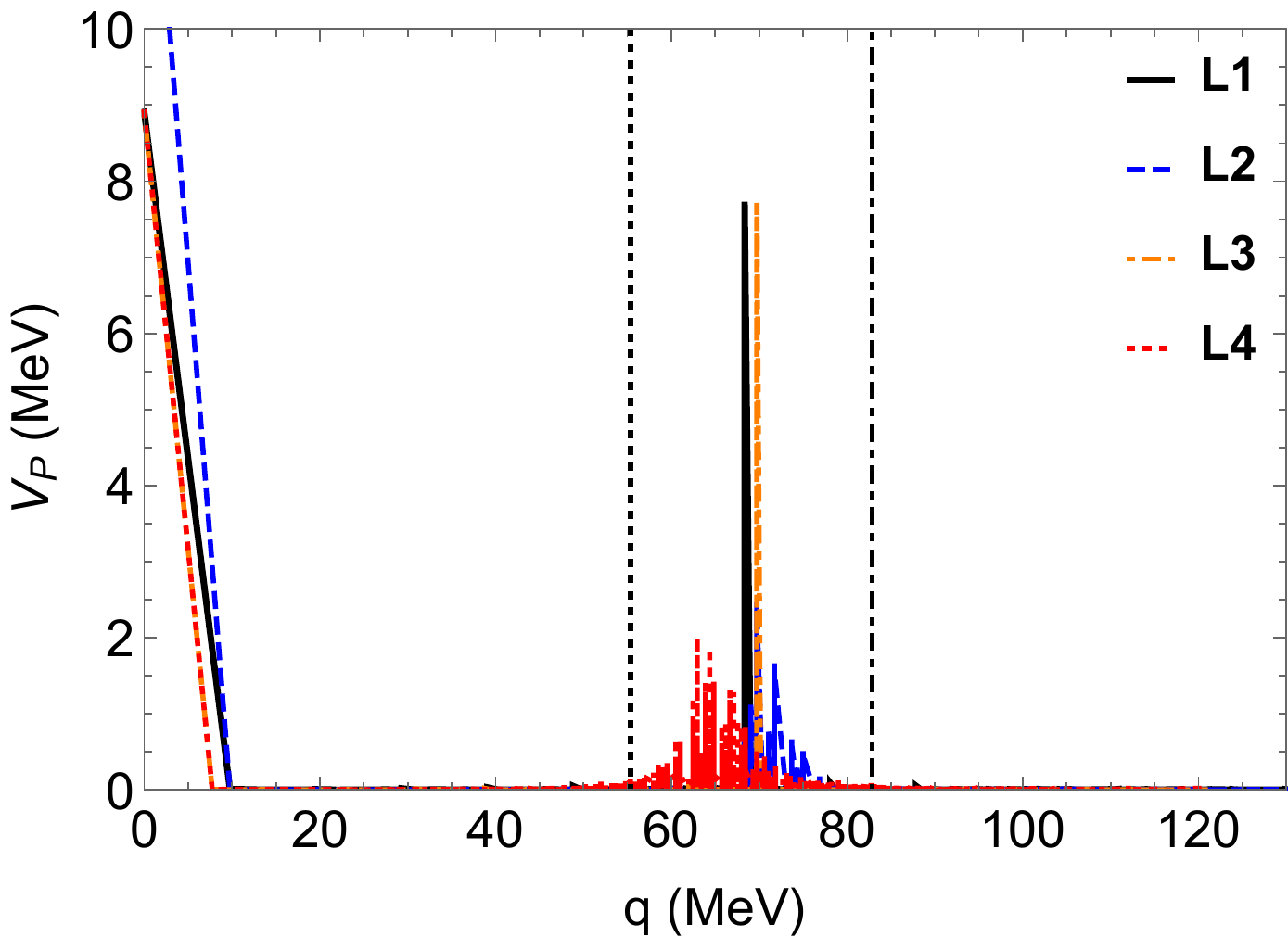} 
	\includegraphics[width=0.45\textwidth]{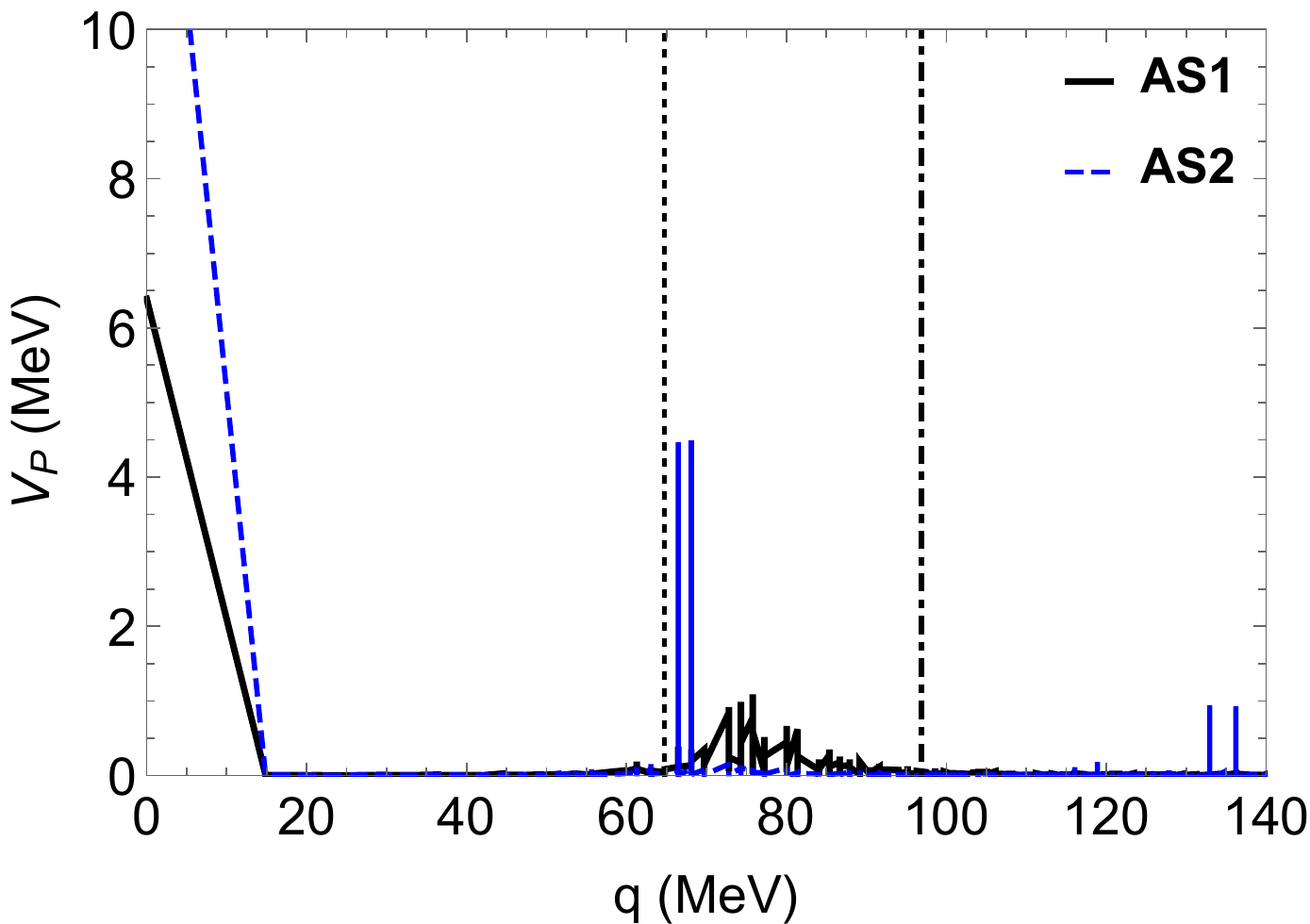} 

	\caption{ Fourier transformed potential energy distribution of a proton in gnocchi (upper left), waffle (upper right), lasagna (lower left), and anti-spaghetti (lower right) respectively. The vertical dot dashed and dotted black lines are lower bounds of allowed $q$ when $Y_e=0.03$ and $Y_e=0.05$ respectively, above which $V(\textbf{q})$ is included in function $R$ (see eq. \ref{eq:R}). The upper bounds of $q$ lie beyond the range of the plots and are not shown here. Properties of these pasta phases are summarized in Table. \ref{tab:md}.  
	}
	\label{fig:vpq}
\end{figure*}

\begin{figure*}[htp]
	\centering
	\includegraphics[width=0.45\textwidth]{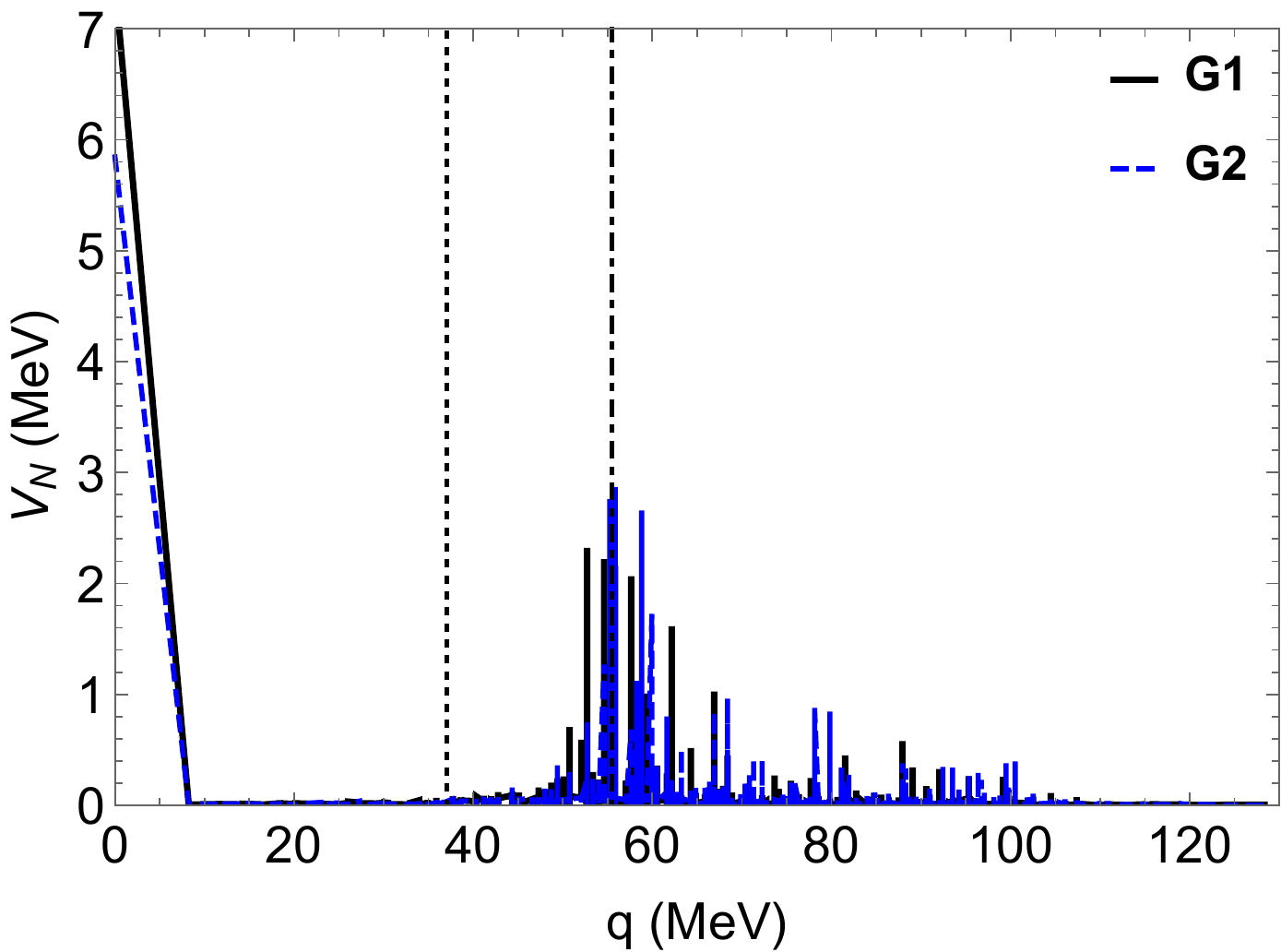}
	\includegraphics[width=0.45\textwidth]{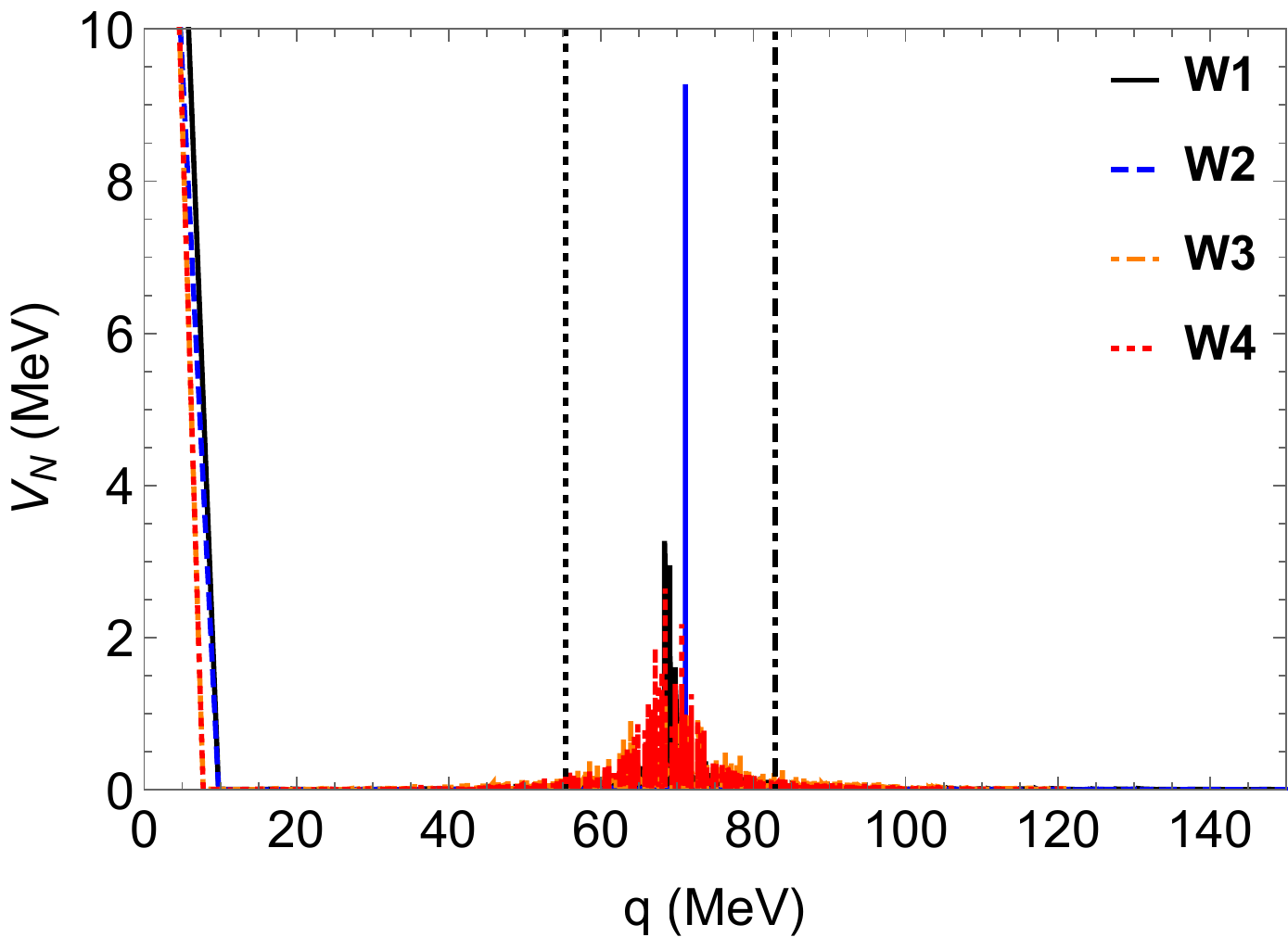}
	\includegraphics[width=0.45\textwidth]{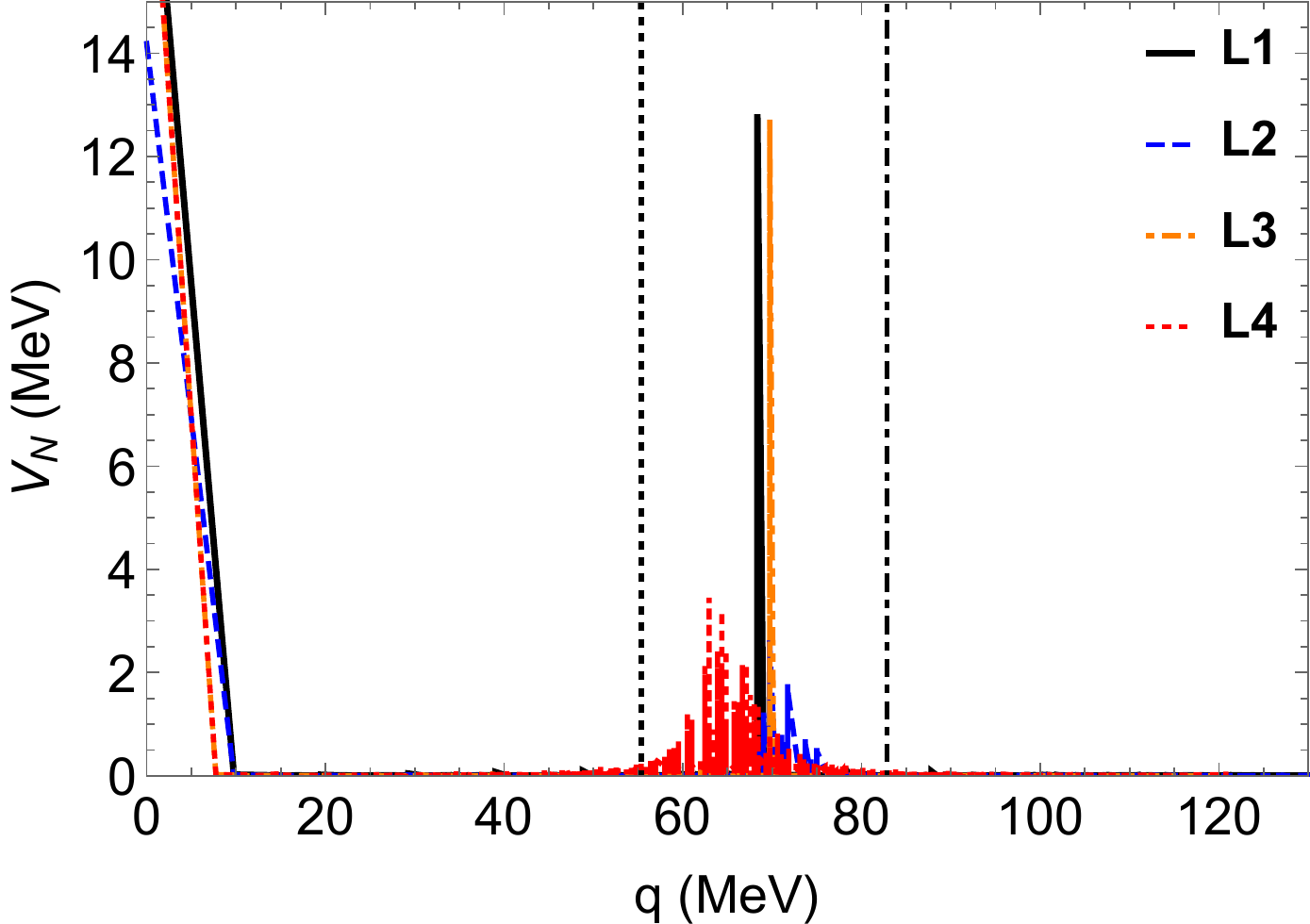} 
	\includegraphics[width=0.45\textwidth]{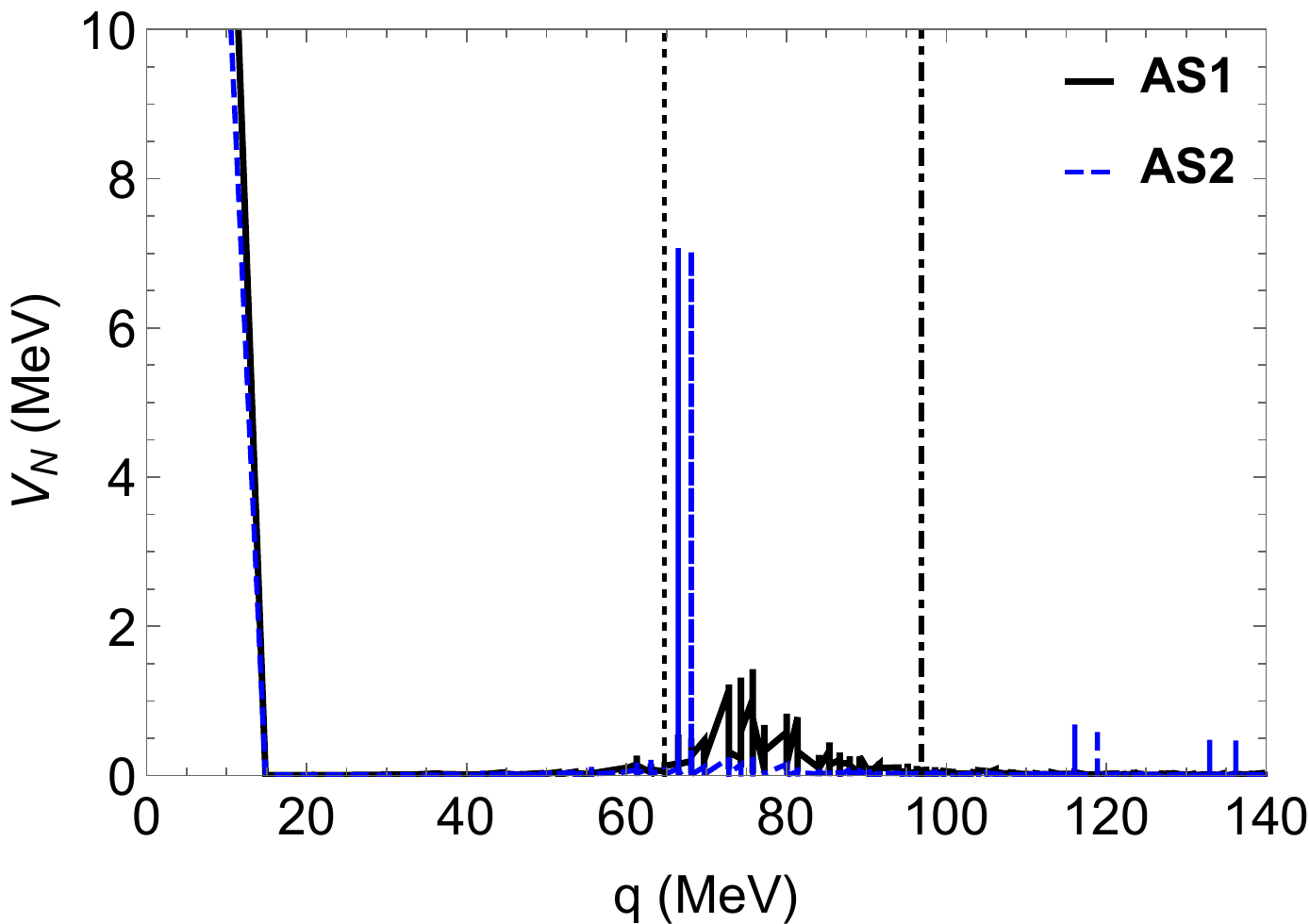} 
	
	\caption{ Fourier transformed potential energy distribution of a neutron in gnocchi (upper left), waffle (upper right), lasagna (lower left), and anti-spaghetti (lower right) respectively. The vertical dot dashed and dotted black lines are lower bounds of allowed $q$ when $Y_e=0.03$ and $Y_e=0.05$ respectively, above which $V(\textbf{q})$ is included in function $R$ (see eq. \ref{eq:R}). The upper bounds of $q$ lie beyond the range of the plots and are not shown here. Properties of these pasta phases are summarized in Table. \ref{tab:md}.  
	}
	\label{fig:vnq}
\end{figure*}

\begin{figure*}[htp]
	\centering
	\includegraphics[width=0.45\textwidth]{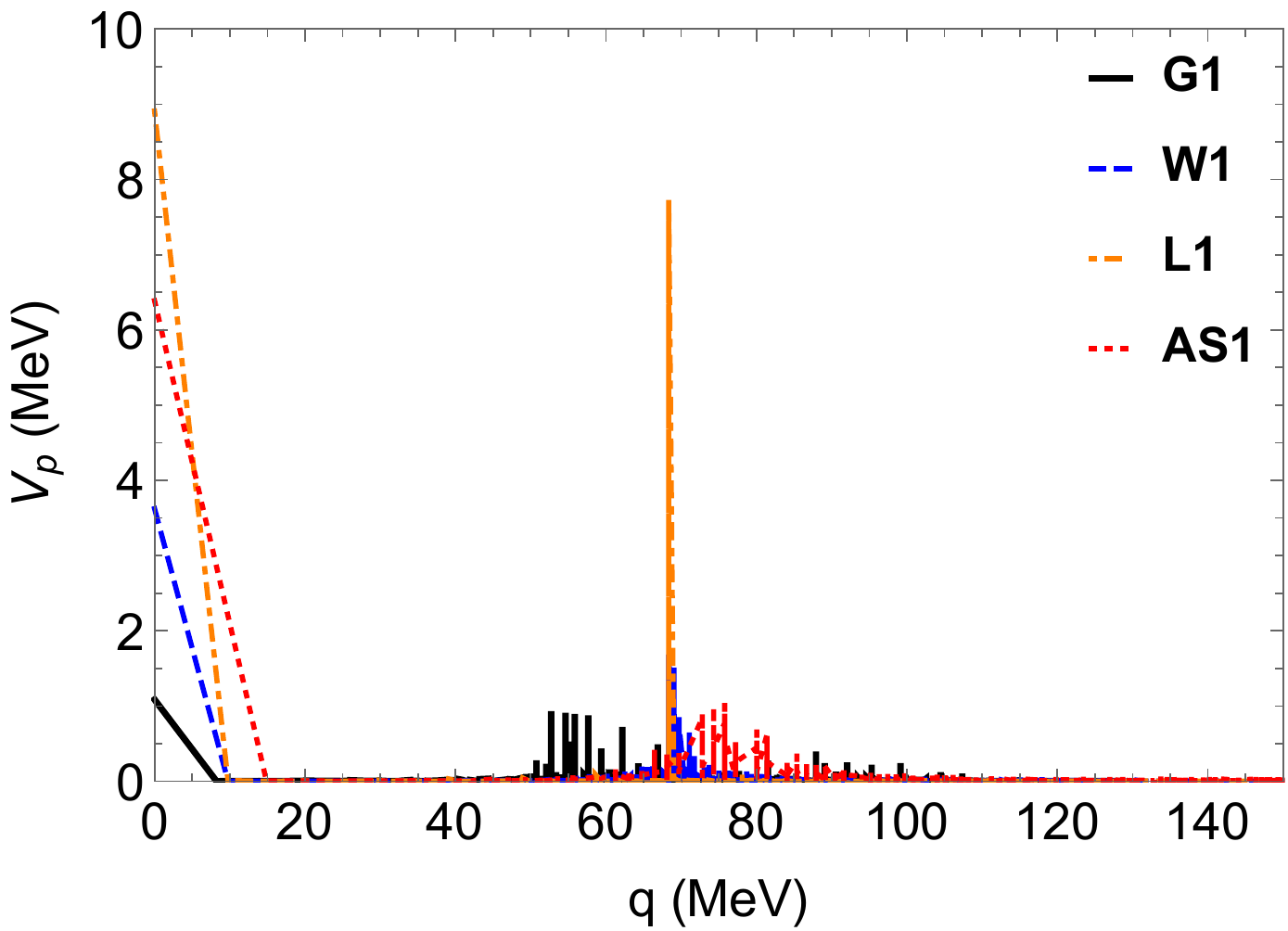}
	\includegraphics[width=0.45\textwidth]{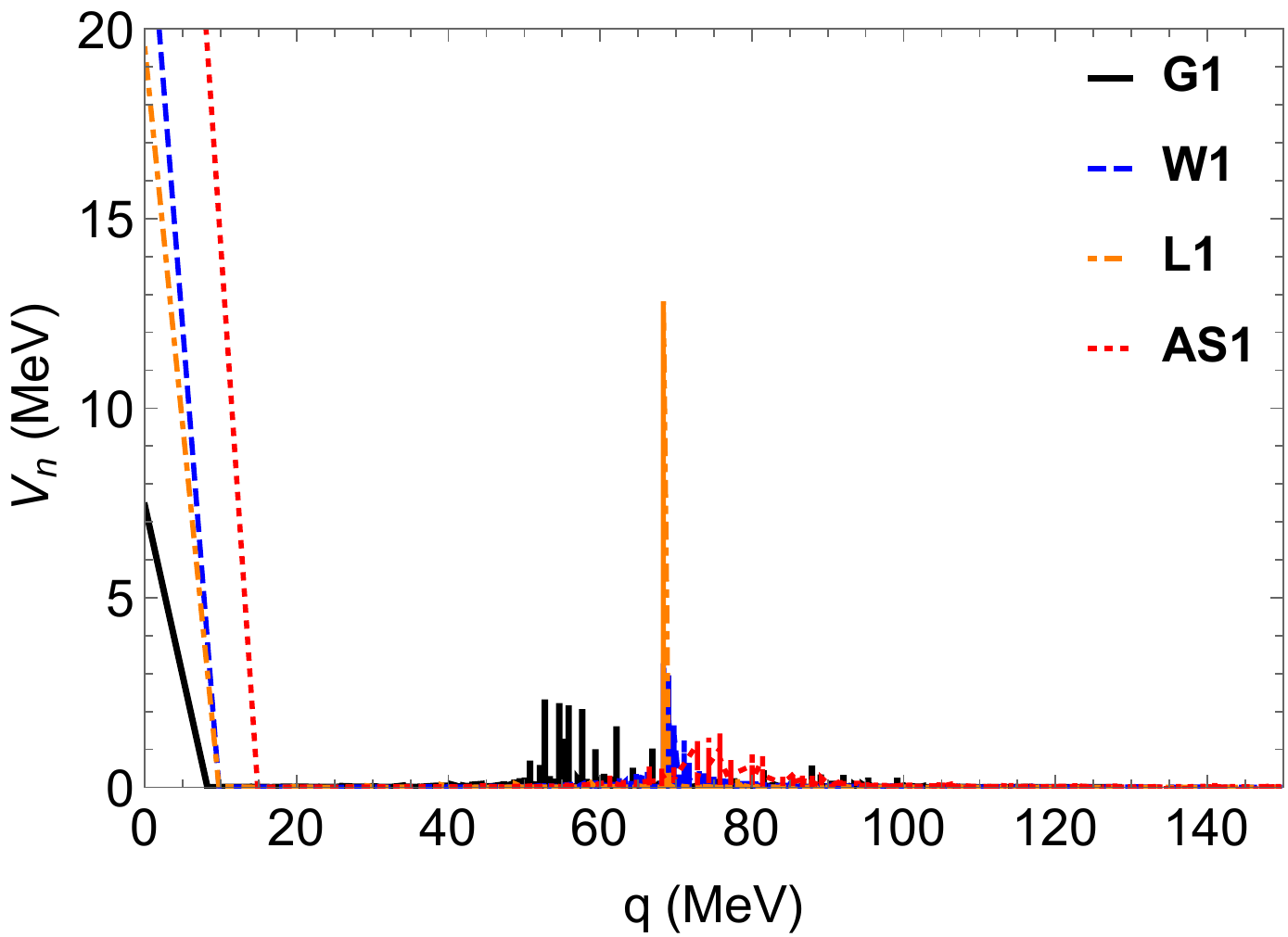}

	\caption{Fourier transformed potential energy distribution of a neutron and a proton in gnocchi, waffle, lasagna and anti-spaghetti are compared. The properties of these pasta phases are summarized in Table. \ref{tab:md}. }
	\label{fig:vqall}
\end{figure*}

We show Fourier transformed proton and neutron potentials as a function of momentum transfer $q$ in Fig. \ref{fig:vpq} and Fig. \ref{fig:vnq}. These potentials for different pasta phases are then compared in Fig. \ref{fig:vqall}. As shown in Fig. \ref{fig:vqall}, the Fourier transformed nucleon potentials display a large peak at  $|\textbf{q}| \simeq 50-80$ MeV 
due to the fact that the periodic spacing of nuclear pasta potential is comparable to the wavelength of the nucleon momentum transfer $\textbf{q}$. To have a clearer understanding of the relationship between the pasta structure and the properties of the peaks of $V(\textbf{q})$, in Appendix A we analytically evaluated the position and the height of the peak of a gnocchi phase assuming it is composed of perfectly spherical nuclei and has a clear BCC lattice structure. We further discuss the possible relationship between the $V(\textbf{q})$ and the static structure factor $S(\textbf{q})$ of the nuclear pasta, where the latter embodies the coherence effect on nuclear pasta electron scattering and on the nuclear pasta neutrino scattering in NSs \cite{Schneider:2013dwa,Horowitz:2016fpa}. The static structure factor of nuclear pasta displays large peaks in $\textbf{q}$ domain when the wavelength of $\textbf{q}$ is comparable to the inter-particle spacing. Due to the structural similarities between distributions of nucleon densities and distributions of nucleon potentials in nuclear pasta (see Figs. \ref{fig:vrball}--\ref{fig:vrasp}, where the neutron density and potential distributions are shown), the peaks of $V(\textbf{q})$ and the peaks of nuclear pasta static structure factor \cite{Horowitz:2016fpa,Horowitz:2004pv} are approximately in the same region of $|\textbf{q}|$. 

Let us now discuss the relationship between the nuclear pasta potentials in real space and the corresponding Fourier transformed potentials. Firstly, in Fig. \ref{fig:vrball} the gnocchi phases G1 and G2 are simulated at different electron fractions, namely $Y_e=0.3$ and $Y_e=0.4$. However the distributions of nucleon potential in real space are very similar to each other , and correspondingly the peaks of $V(\textbf{q})$ of these two simulations approximately overlap with each other, as shown in the upper left panel of Fig. \ref{fig:vpq} and \ref{fig:vnq}. Secondly, the lasagna phases L1-4 is simulated with different number of nucleons, different $Y_e$ and orientations of the lasagna plates, as shown in Fig. \ref{fig:vrlasagna}. In the left lower panel of Fig. \ref{fig:vpq} and Fig. \ref{fig:vnq}, the peak of $V(\textbf{q})$ based on simulation L1 looks very similar to that based on L3, which indicates that the number of nucleons involved in our simulations will not severely affect the outcome, and that the finite-size effect of our MD simulations is minor. However, although the location of the peaks based on these four simulations basically agree with each other, the height of peaks based on L2 and L4 is obviously smaller than those based on L1 and L3. This is due to the fact that lasagna simulations of L2 and L4 exhibit more irregular local structures such as the connection between two plates and the curvature of the plates, while keeping about the same spacing of plates as in L1 and L3. Thirdly, the waffle phase are simulated with different number of nucleons, electron fractions $Y_e$, and orientations of the waffle plates. In the upper right panel of Fig. \ref{fig:vpq} and Fig. \ref{fig:vnq}, the distribution of $V(\textbf{q})$ based on W1 and W3 are similar, which once again demonstrate the finite size effect on our evaluations is small. But the peak of W1 is obviously higher than the other three simulations, which is possibly due to a more regular distribution of these waffle plates and short-lived holes in this specific simulation, as shown in Fig. \ref{fig:vrwaffle}. Finally, we present the distribution of $V(\textbf{q})$ corresponding to anti-spaghetti in the lower right panel in Fig. \ref{fig:vpq} and Fig. \ref{fig:vnq}. Although the location of the main peaks based on AS1 and AS2 are approximately the same, the height of peaks based on these two simulations are very different. This is because AS1 is actually a disordered form of AS2, and the latter shows clearly the long-range correlations that AS1 lacks and exhibits a much clearer periodic structure than AS2 does, as shown in Fig. \ref{fig:vrasp}.

\subsection{Neutrino emissivity}\label{result:emissivity}

\begin{table*}[t] \caption{Summary of Direct Urca emissivity in this work. We include function $R$, neutrino emissivity $Q$ (in unit of $10^{21} \text{ erg cm}^{-3}\text{s}^{-1}$  and the ratio of neutrino luminosity $L$ via direct Urca from the crust at $Y_e=0.03$ and at $Y_e=0.05$ to that via modified Urca from the core. 
The temperature corresponding to the calculations here is  $T=3\times 10^8$ K.  }\label{tab:emission}
\begin{ruledtabular}
\begin{tabular}{lllllll}
Identifier &$R_{Y_e=0.03}$&$R_{Y_e=0.05}$& $Q^{crust}_{Y_e=0.03}$($10^{21} \text{ erg cm}^{-3}\text{s}^{-1}$) &$Q^{crust}_{Y_e=0.05}$ ($10^{21} \text{ erg cm}^{-3}\text{s}^{-1}$)&$L^{crust}_{Y_e=0.03}/L^{core}$& $L^{crust}_{Y_e=0.05}/L^{core}$ \\
\hline
G1 &0.046&0.58& $19$ & 280&$1.9\times10^3$&$2.9\times10^{4}$ \\
G2 &0.11&0.95& $44$ & 470&$4.5\times10^3$&$4.8\times10^{4}$  \\
L1 &$9.2\times10^{-5}$&0.81&$5.7\times10^{-2}$ & 590&5.8& $6.1\times10^{4}$ \\
L2 &$1.2\times10^{-4}$&0.27& $7.3\times10^{-2}$ & 200&7.5& $2.0\times10^4$ \\
L3 &$1.6\times10^{-5}$&0.85& $9.8\times10^{-3}$ & 620&1.0& $6.4\times10^4$ \\
L4 &$2.7\times10^{-4}$&0.36& $0.17$ & 260&17&$2.7\times10^{4}$  \\
W1 &$5.6\times10^{-4}$&0.16& $0.35$ & 120&36& $1.2\times10^4$ \\
W2 &$1.8\times10^{-4}$&0.27& $0.11$ & 190&12&$2.0\times10^{4}$  \\
W3 &$9.1\times10^{-4}$&0.14& $0.56$ & 100&58& $1.0\times10^4$ \\
W4 &$4.9\times10^{-4}$&0.26& $0.3$ & $190$&31&$2.0\times10^4$\\
AS1 &$7.6\times10^{-4}$&0.047& $0.57$ & $41$&58&$4.2\times10^3$\\
AS2 &$5.9\times10^{-3}$&0.03& $4.4$ & $27$&452&$2.8\times10^3$ \\
\end{tabular}
\end{ruledtabular}
\end{table*}

\begin{figure*}[htp]
	\centering
	\includegraphics[width=0.45\textwidth]{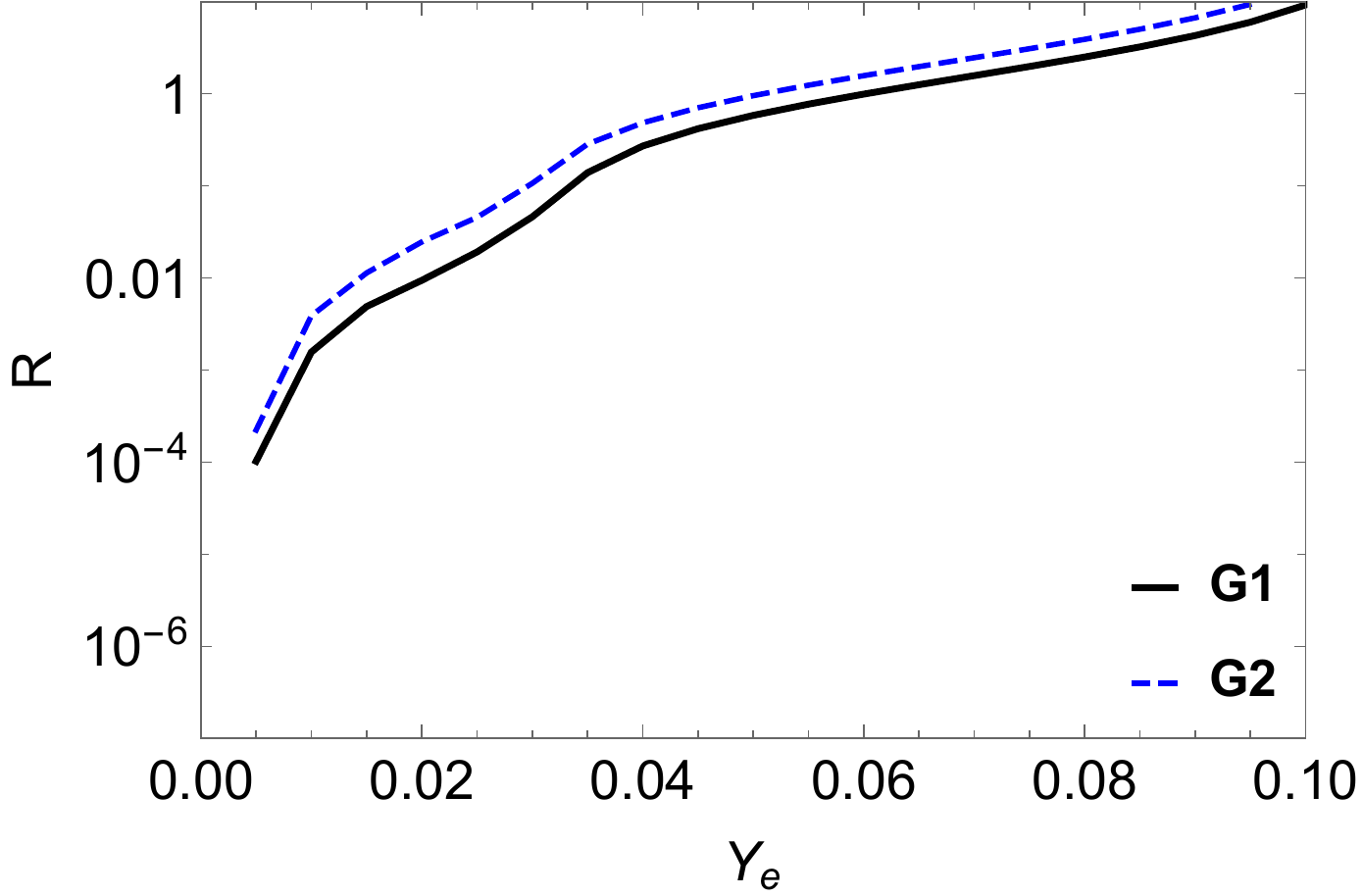}
	\includegraphics[width=0.45\textwidth]{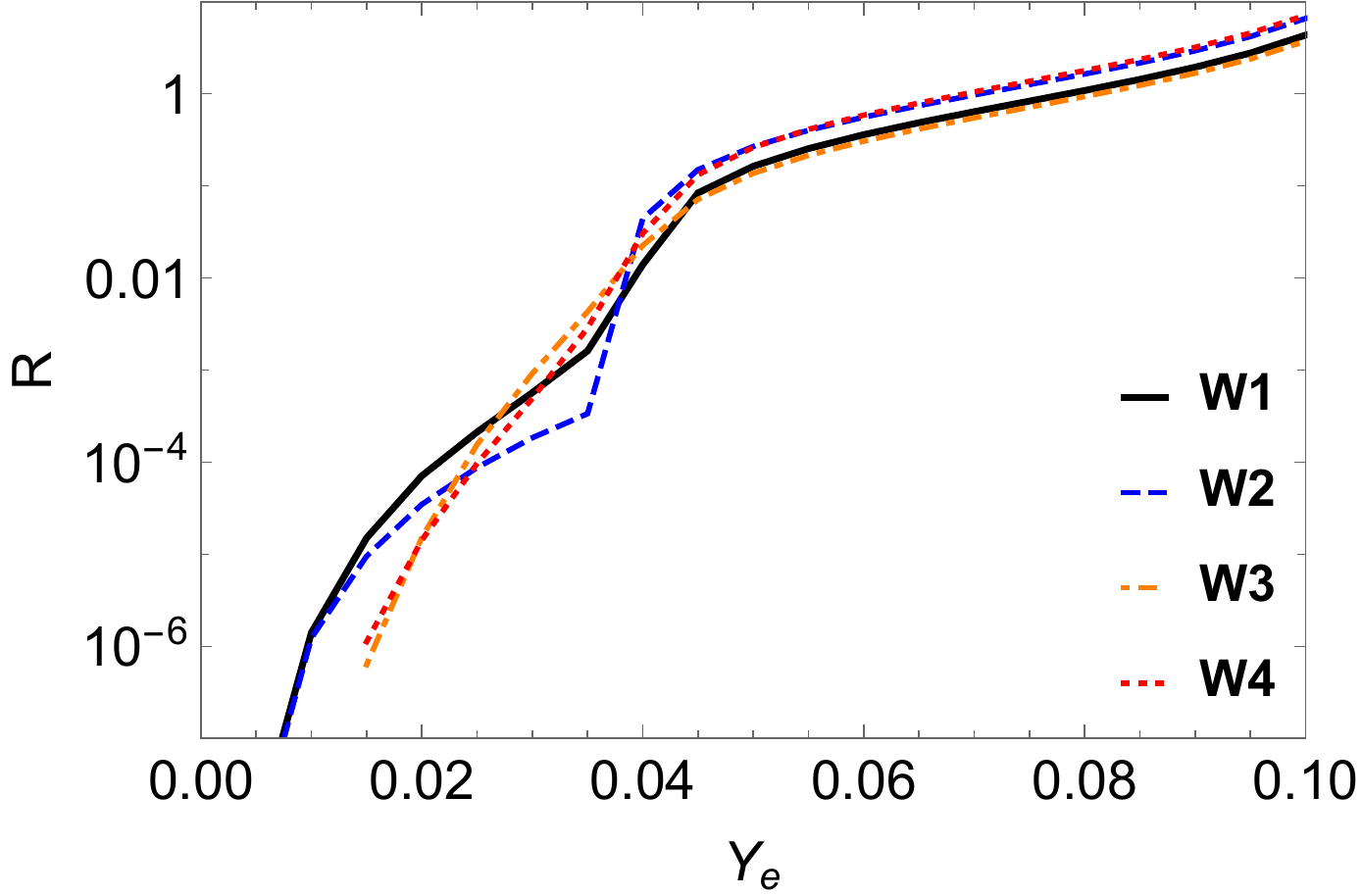} 
	\includegraphics[width=0.45\textwidth]{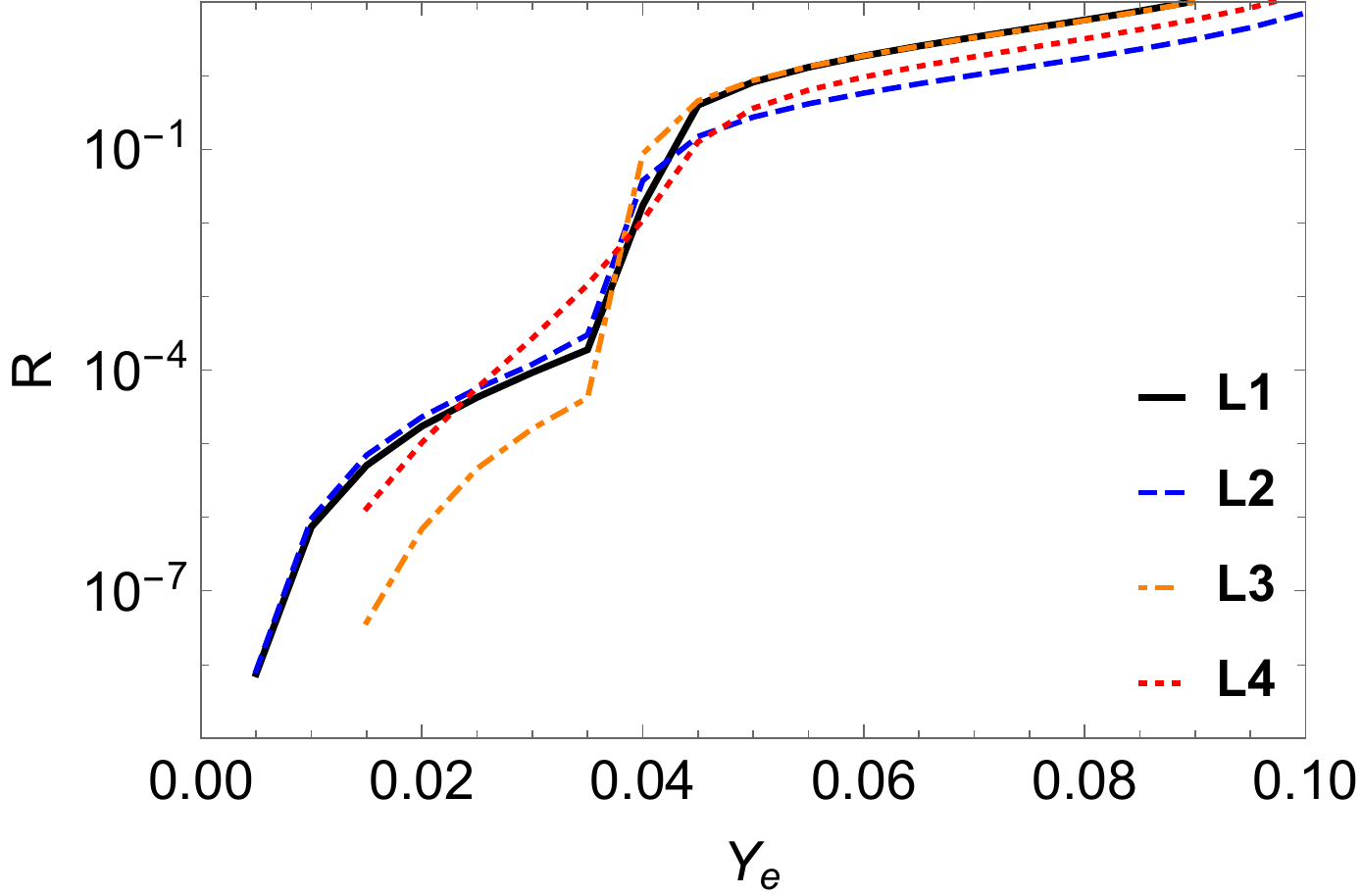} 
	\includegraphics[width=0.45\textwidth]{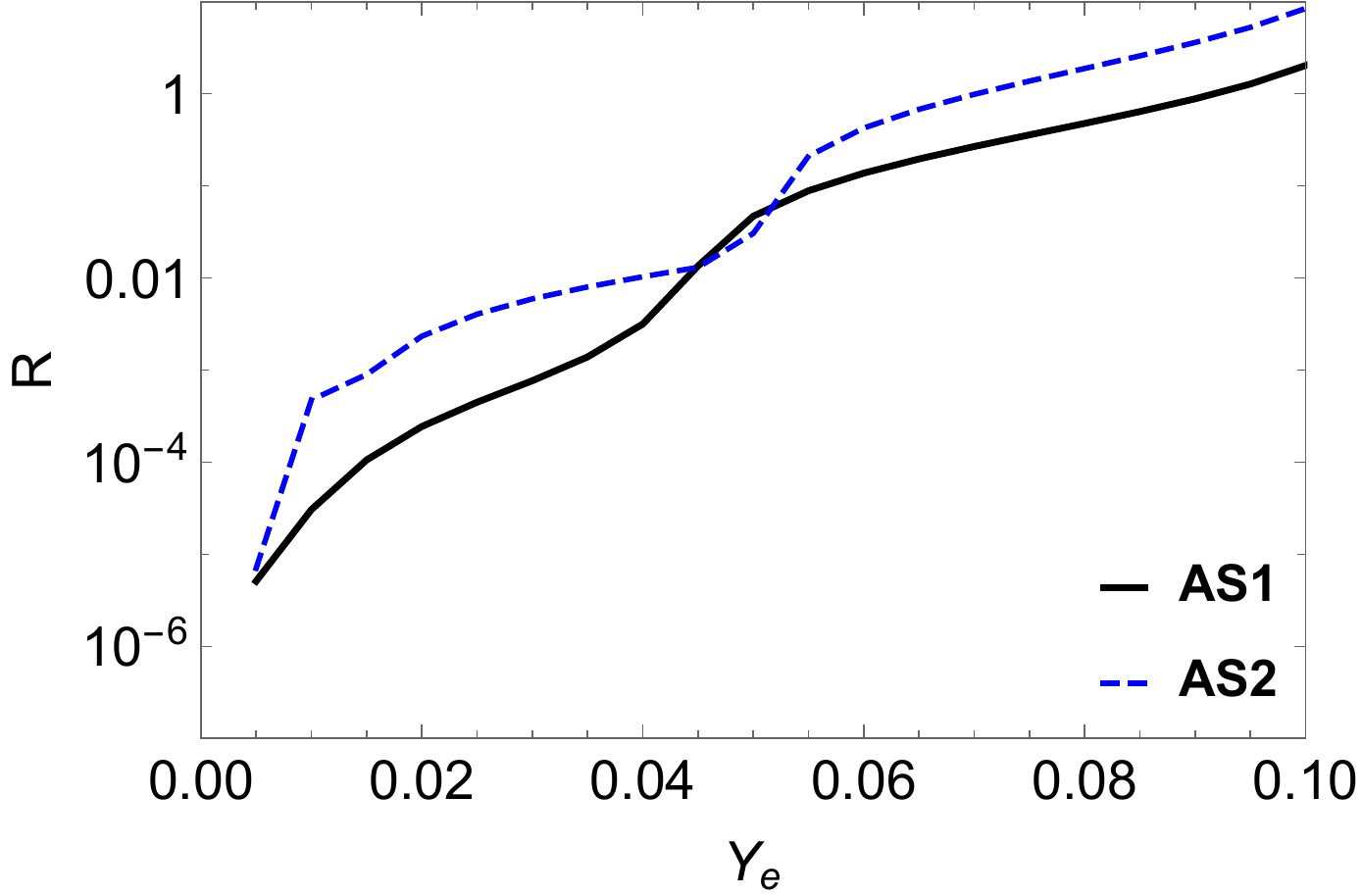} 
	\caption{ The factor $R$ as a function of $Y_e$ in gnocchi (upper left), waffle (upper right), lasagna (lower left) and anti-spaghetti (lower right).}
	\label{fig:RYe}
\end{figure*}

\begin{figure*}[htp]
	\centering
	\includegraphics[width=0.45\textwidth]{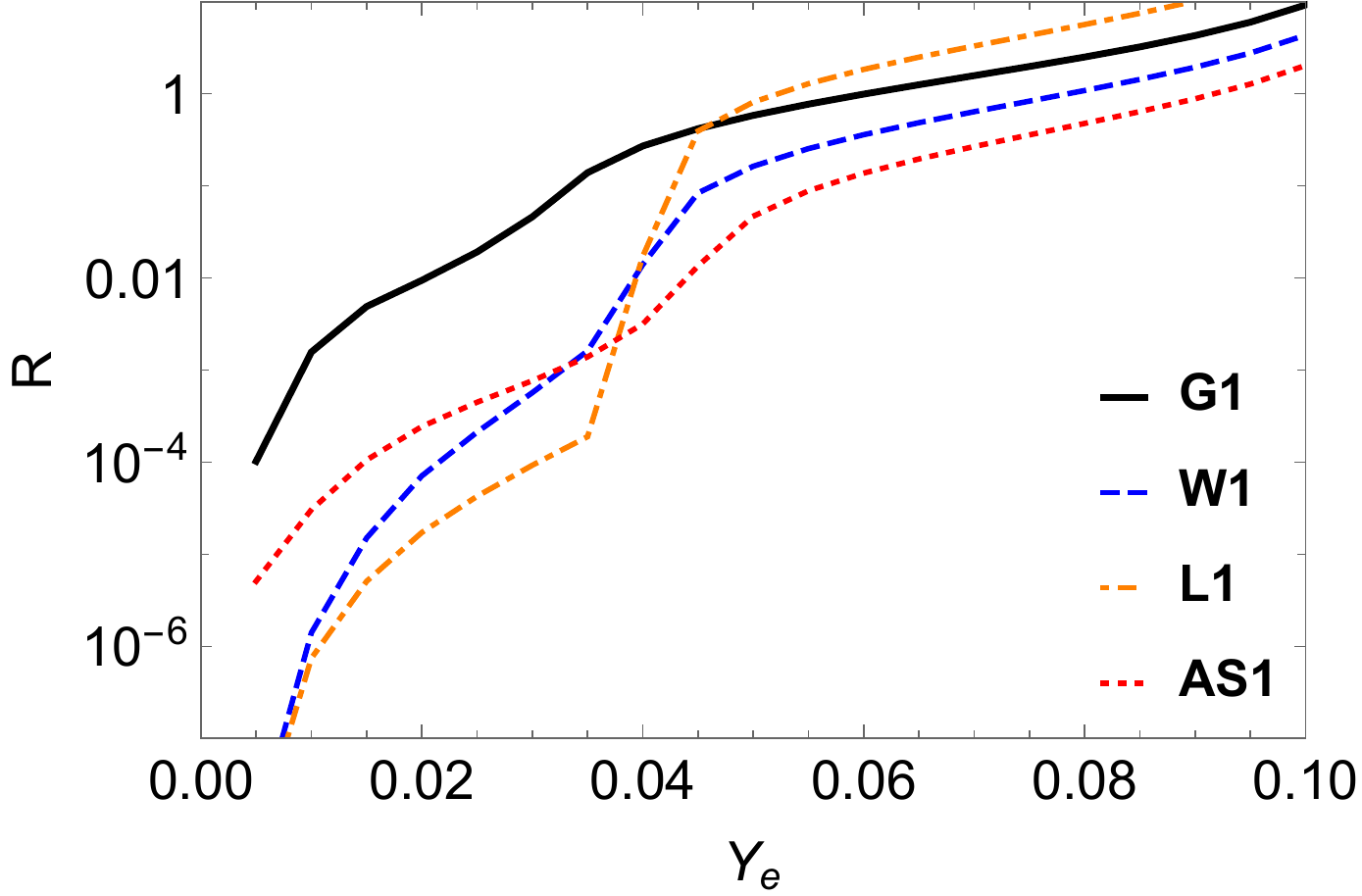}
	\includegraphics[width=0.45\textwidth]{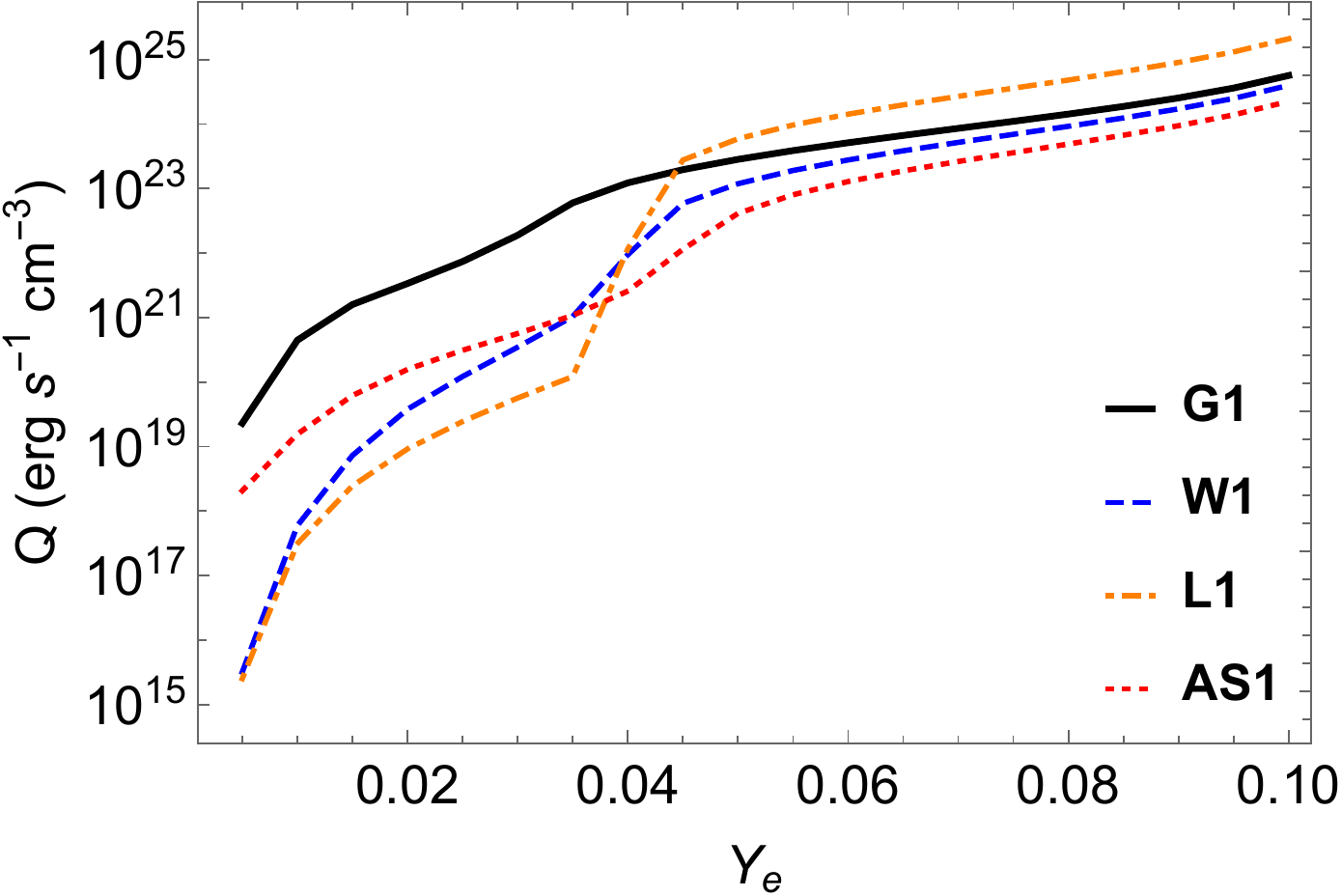}
	\caption{ The left panel shows $R$ as a function of $Y_e$ in different nuclear pasta phases, which are G1, W1, L1 and AS1 respectively. The right panel shows neutrino emissivities in these phases. The corresponding densities of G1, L1, W1 and AS1 pasta phases are listed in Table 1. The temperature at which Q is calculated is at $T=3\times10^8~K$.    }
	\label{fig:RYeAll}
\end{figure*}

In this section we calculate the direct Urca neutrino emissivity in nuclear pasta. The effect of nuclear pasta structure on the neutrino emissivity is illustrated in Fig. \ref{fig:RYe} and Fig. \ref{fig:RYeAll}, where $R$ varies as a function of $Y_e$, at fixed baryon densities. As $Y_e$ decreases, the lower bound of allowed $\textbf{q}$ rises up, and the contribution from the peaks of $V(\textbf{q})$ to the $R$ will be excluded if the lower bound is higher than $\textbf{q}_{\mathrm{peak}}$, with $\textbf{q}_{\mathrm{peak}}$ being the corresponding momentum transfer at the peak of $V(\textbf{q})$.  As a result, the function $R$ decreases rapidly at around $Y_e=$0.01, 0.035, 0.035 and 0.045 for gnocchi, waffle, lasagna and anti-spaghetti phase respectively in Fig. \ref{fig:RYe} and Fig. \ref{fig:RYeAll}. We note that the electron fractions for nuclear pasta in inner crust at beta equilibrium is approximately $0.03<Y_e<0.035$, based on the liquid drop models \cite{Oyamatsu:1993zz}.  Indeed, the region of $Y_e$ at beta equilibrium in inner NS crust demonstrate the close proximity to the enhancement of $R$ and hence the enhancement of neutrino emissivities via direct Urca reactions due to nuclear pasta structures. In Tab. \ref{tab:emission} we summarize the $R$s of different pasta phases. To illustrate the contribution from $V(\textbf{q})$ peaks, $R$ is calculated in Table \ref{tab:emission} at two different electron fractions, $Y_e=0.03$ and $Y_e=0.05$. As shown in Fig. \ref{fig:vpq} and Fig. \ref{fig:vnq}, at $Y_e=0.03$ the $R$ of most nuclear pasta phases (except those corresponding to G1 and G2) do not include the contributions from $V(\textbf{q})$ peaks because of momentum conservation. At $Y_e=0.05$, all pasta phases summarized in this table have large $R$ due to the contribution from peaks of their $V(\textbf{q})$. Correspondingly, the neutrino emissivity is greatly enhanced, and is only 1-2 orders of magnitude weaker than that via direct Urca reactions. However, when the peaks of $V(\textbf{q})$ are excluded due to momentum conservation requirement, the $R$ decreases by 3-4 magnitude of orders . In the latter case our results about $R$ are reasonably close to the calculations in \cite{Gusakov:2004mj}, while the remaining deviations between our results of $R$ and that reported in \cite{Gusakov:2004mj} are possibly due to the differences between the pasta models applied. 

Finally, the neutrino emissivity $Q$ from the core via modified Urca reactions are compared to $Q$ from the nuclear pasta layer in the inner crust due to direct Urca reactions. The neutrino emissivities  of the modified Urca process, in both the neutron and proton branches (which is denoted as $Q_{MN}$ and $Q_{MP}$ respectively), are summarized  below (see detailed description of modified Urca in \cite{Yakovlev:2000jp}): 
\begin{equation}
    Q_{MN}\approx8.1\times10^{21}\left(\dfrac{n_P}{n_0}\right)^{1/3}T_9^8\alpha_n\beta_n\;\mathrm{ erg\,cm^{-3}s^{-1}},
\end{equation}
\begin{equation}
    Q_{MP}\approx Q_{MN}\dfrac{(P_{Fe}+3P_{FP}-P_{FN})^2}{8P_{Fe}P_{FP}}\Theta_{MP},
\end{equation}
 where $\alpha_n=1.13$, $\beta_n=0.68$, and $\Theta_{MP}$ is the threshold for the proton branch, allowing the modified Urca process when $P_{FN}<4P_{Fe}$. We calculate $Q_{MN}$ and $Q_{MP}$ at core density $n_{\mathrm{core}}=2n_0$, where $n_0$ is the saturation density $0.16\, fm^{-3}$. The neutron star is assumed to be isothermal and neutrino emissivity from the core and the crust are both calculated at $T=3\times10^8$ K.

Given $Q_{MN}$ and $Q_{MP}$, in Tab. \ref{tab:emission} we list  an order-of-magnitude estimate on the ratio of crust neutrino luminosity to core neutrino luminosity at $Y_e=0.03$ and $Y_e=0.05$. The neutrino luminosity of modified Urca from the core is:
\begin{equation}
    L^{\mathrm{core}}\approx\dfrac{4\pi}{3}R_{\mathrm{NScore}}^3(Q_{MN}+Q_{MP}),
\end{equation}
where $R_{\mathrm{NScore}}=10$ km is approximately the radius of neutron star cores. The neutrino luminosity of direct Urca in the crust is:
\begin{equation}
    L^{\mathrm{crust}}\approx4\pi R_{\mathrm{NScore}}^2hQ,
\end{equation}
where $h=100$ m is approximately the width of nuclear pasta layer \cite{Stone:2009rp}. In the calculations of $L^{\mathrm{crust}}$, we assume that the inner crust is composed of nuclear pasta of a specific phase, \emph{e.g.}, only lasagna or only anti-spaghetti. A more accurate estimation of $L^{\mathrm{crust}}$ might require considering the co-existence of multi-nuclear pasta phases in the inner crust, so that the total luminosity would be an appropriate average of the luminosities of the various pasta phases. In Tab. \ref{tab:emission} we summarize the neutrino luminosities of direct Urca process due to different nuclear pasta phases in neutron star crusts, and see that they can be about 1-2 magnitude of orders stronger than that in neutron star cores from modified Urca process, if the contribution from peaks of $V(\textbf{q})$ to neutrino emissivity $Q$ is excluded by momentum conservation. At sufficiently high $Y_e$ (for example at $Y_e=0.05$), contributions from the peaks of $V(\textbf{q})$ to the function $R$ can greatly amplify the neutrino luminosity in neutron star inner crusts, making it even stronger, which is about 3-4 magnitude of orders higher than that due to the modified Urca reactions in the cores of NSs.  

\section{Conclusion}\label{conclude}

In this paper we calculated the neutrino emissivity due to a direct Urca process in nuclear pasta.  This nonuniform phase is expected near the base of the neutron star crust.
Different shaped pasta phases were explored using molecular dynamics simulations containing 51,200 and 204,800 nucleons. In our semi-classical simulations, both neutrons and protons are free to explore a variety of shapes.  We approximated the wave functions of nucleons in our pasta simulations using perturbation theory as in Ref. \cite{Gusakov:2004mj}. Given these nucleon wave functions, the neutrino emissivity of the direct Urca process was calculated for various nuclear pasta phases, including gnocchi, waffle, lasagna, and anti-spaghetti. 
We found that the neutrino luminosity due to a direct Urca process in nuclear pasta can be 3-4 orders of magnitude larger than that from the modified Urca process in neutron star cores.  Thus neutrino radiation from pasta could dominate over radiation from the core.  This enhanced emission could have a pronounced effect on the cooling of neutron stars and on X-ray observations of NS thermal radiations.
Therefore, future work should explore further the neutrino emissivity of nuclear pasta including that from fully quantum mean field calculations.  This will allow calculations directly at low beta equilibrium values of $Y_e$ and should provide a better understanding on how neutrino emissivitiy depends on $Y_e$. 

In the near future we expect more X-ray observations of thermal radiation from NS.  These cooling observations may depend on a variety of NS features such as the presence or absence of a heavy element envelope, of a Direct Urca process in the core, and on a variety of superfluid and superconducting pairing gaps \cite{Page_2004}. In addition there could be a sizable contribution to cooling from nuclear pasta. The pasta mechanism, if effective, should be present in all neutron stars and this will, if all other NS features remain unchanged, make all stars cool faster and therefore reach lower temperatures over a given time interval. Therefore, the existence of nuclear pasta and its associated direct Urca emissions, together with other NS features mentioned above, should all be taken into account as possible cooling variables in the future work of NS cooling simulations. It should be possible to use X-ray observations of both isolated and accreting NS to sort out some features of NS cooling and gain insight into the dense phases of matter present in NS.    


\subsection*{Acknowledgments}

ZL thanks D. G. Yakovlev and S. P. Harris for helpful discussions. ZL and CL acknowledge funding from the National Science Foundation Grant No. PHY-1613708. ZL acknowledges funding from US Department of Energy grants DE-SC0019470. CH is supported in part by US Department of Energy grants DE-FG02-87ER40365 and DE-SC0018083. This research was supported in part by Lilly Endowment, Inc., through its support for the Indiana University Pervasive Technology Institute, and in part by the Indiana METACyt Initiative. The Indiana METACyt Initiative
 at IU was also supported in part by Lilly Endowment, Inc. This material is based upon work supported by the National Science Foundation under Grant No. CNS-0521433. T.


\appendix
\section{Analytic model of nucleon potentials for the gnocchi phase}

In the appendix we aim to gain an analytical understanding of the large-scale molecular dynamics numerical simulations. We choose the gnocchi phase as the test bed, which forms a body-centered-cubic (BCC) lattice when the simulation is equilibrated. The Fourier transformed potential $V_j(\textbf{q})$ is defined as: 
\begin{equation}\label{vqana}
    V_j(\textbf{q})=\dfrac{1}{V}\int_VdVV_j(\textbf{r})e^{i\textbf{q}\cdot\textbf{r}}.
\end{equation}
In a well equilibrated gnocchi phase a reciprocal lattice structure is formed and we assume that $V_j(\textbf{r})=V_j(\textbf{r}+\textbf{T})$, where $\textbf{T}$ is the lattice vector. $V_j(\textbf{r})$ can be expressed in terms of a Fourier decomposition, given its periodic structure
\begin{equation}\label{vrfourier}
    V_j(\textbf{r})=\sum_{\textbf{G}}V_Ge^{-i\textbf{G}\cdot\textbf{r}},
\end{equation}
where $\textbf{G}$ is the reciprocal lattice vector.
Given eq. \ref{vrfourier}, eq. \ref{vqana} becomes
\begin{equation}
        V_j(\textbf{q})=\sum_G\int_VdVV_Ge^{i(-\textbf{G}+\textbf{q})\cdot\textbf{r}}.
\end{equation}
We see that $V(\textbf{q})$ reaches its peak at the diffraction points where $\textbf{q}=\textbf{G}$, and find:
\begin{equation}
    \begin{split}
        V_j(\textbf{G})&=\dfrac{1}{V}\sum_T\int_{\mathrm{cell}}dVV_j(\textbf{r}+\textbf{T})e^{i\textbf{G}\cdot(\textbf{r}+\textbf{T})}\\&=\dfrac{N}{V}\int_{\mathrm{cell}}dVV_j(\textbf{r})e^{i\textbf{G}\cdot\textbf{r}},
    \end{split}
\end{equation}
where $N$ is the number of unit lattice cells in a MD simulations and $V$ is the volume of the box in MD simulations. Assuming that we have s gnocchi in a unit cell located at $\textbf{r}_k$, it is convenient to write potential energy as the superposition of potential energy $V_j$ associated with each gnocchi $k$ of the basis, so that $V_j(\textbf{r})=\sum_{k=1}^SV_j(\textbf{r}-\textbf{r}_k))$. We then have
\begin{equation}
\begin{split}
    \int_{cell}dV V_j(\textbf{r})e^{-i\textbf{G}\cdot\textbf{r}}&=\sum_{k=1}^S\int_{\mathrm{cell}}dV V_j(\textbf{r}-\textbf{r}_k)e^{-i\textbf{G}\cdot\textbf{r}}\\&=\int_{\mathrm{cell}}dVV_j(\textbf{R})e^{-i\textbf{G}\cdot\textbf{R}}\times\sum_{k=1}^S e^{-i\textbf{G}\cdot\textbf{r}_k}~,
\end{split}
\end{equation}
where $\textbf{R}=\textbf{r}-\textbf{r}_k$.
Assuming that the nucleon potential energy $V_j$ in the gnocchi is distributed uniformly with a sharp surface radius $R$, eq. 9 could be further simplified, since
\begin{equation}
    \int_{\mathrm{cell}}dVV_j(\textbf{R})e^{-i\textbf{G}\\\bf{R}}=4\pi V_j\dfrac{-GR \cos(GR)+\sin(GR)}{G^3}.
\end{equation}
For a BCC lattice, we have $\sum_{k=1}^S e^{-i\textbf{G}\cdot\textbf{r}_j}=1+e^{-i\textbf{G}\textbf{r}_1}$, where $\textbf{G}=\dfrac{2\pi}{a}(m_1\hat{i}+m_2\hat{j}+m_3\hat{k})$ and $\textbf{r}_1=\dfrac{a}{2}(\hat{i}+\hat{j}+\hat{k})$, with $a$ being the center-to-center distance of the BCC lattice. It turns out that $\sum_{k=1}^S e^{-i\textbf{G}\cdot\textbf{r}_j}=1+(-1)^{m_1+m_2+m_2}$, and we have
\begin{equation}
\begin{split}
     V_N(\textbf{G})=4\pi V_j\dfrac{-GR \mathrm{cos}(GR)+\mathrm{sin}(GR)}{G^3}\\\times[1+(-1)^{m_1+m_2+m_2}]\dfrac{N}{V}.
\end{split}
\end{equation}
Finally, we compare the analytical expression of $V_N(\textbf{G})$ with that from numerical simulations. In the gnocchi phase G1, the gnocchi center-to-center distance is approximately 30 fm, the radius of the sphere is approximately 7.5 fm, and the mean potential energy of neutrons in gnocchi is approximately 30 MeV. Plugging these numbers into eq. A7, we found that the first peak locates at $|\textbf{q}|=|\textbf{G}|\approx58$ MeV when $m_1+m_2+m_3=2$, and $V_N(\textbf{G})\approx2.3 MeV$, which agrees with our numerical results as shown in Fig. \ref{fig:vnq} quite well.

\bibliography{b}

   

   

  
  
  

\begin{appendix}

\end{appendix}

\end{document}